\documentclass[10pt,twocolumn]{article}

\usepackage{authblk}
\usepackage{amsmath}
\usepackage{graphicx}
\usepackage{upgreek}
\usepackage{xcite}
\usepackage{color}
\usepackage[margin=2cm]{geometry}
\usepackage[normalem]{ulem}
\usepackage[square,sort&compress,numbers]{natbib}

\title{Harnessing energy landscape exploration to control the buckling of cylindrical shells}
\author[1]{Panter J. R.}
\author[2]{Chen J.}
\author[2]{Zhang T.}
\author[1]{Kusumaatmaja H.}
\affil[1]{Department of Physics, Durham University, South Road, Durham, DH1 3LE, UK}
\affil[2]{Department of Mechanical and Aerospace Engineering, Syracuse University, Syracuse, NY 13244, USA}

\date{}
\begin{document}

\twocolumn[
\begin{@twocolumnfalse}
\maketitle

\begin{abstract}

The complexity and unpredictability of postbuckling responses in even simple thin shells have raised great challenges to emerging technologies exploiting buckling transitions. Here we comprehensively survey the buckling landscapes to show the full complexity of the stable buckling states and the transition mechanisms between each of them. This is achieved by combining a simple and versatile triangulated lattice model for modelling the shell morphologies with efficient high-dimensional free-energy minimisation and transition path finding algorithms. We show how the simple free energy landscapes of short, lightly compressed cylinders become vastly more complex at high compressive strains or aspect ratios. We then exploit these landscapes to introduce an effective method for targeted design – landscape biasing. This is used to inform thickness modifications enabling landscape redesign, and the development of structures which are highly resistant to lateral perturbations. Our methods are general, and can be extended to studying postbuckling responses of other geometries.

\end{abstract}

\end{@twocolumnfalse}
]

\section{Introduction}

Having shed the perception of being a purely problematic phenomenon \cite{Reis2018}, postbuckling responses are rapidly being shown to enable a broad range of emerging technologies \cite{Reis2015,Hu2015,Zhang2017,Bertoldi2017}. Such applications include soft robotics and actuation \cite{Rus2015,Hines2017}, mechanical metamaterials \cite{Yang2019} including origami- and kirigami-inspired designs \cite{Zhai2018}, morphable soft electronics \cite{Fu2018, Ning2018}, logic gates \cite{Song2019},  energy harvesting \cite{Dagdeviren2016}, damping devices \cite{Haghpanah2017}, information storage \cite{Chung2018}, and bioinspired design \cite{Meyers2013}. However, in general the extreme complexity of these responses \cite{Knobloch2015} has largely limited studies to investigate simple structures with very few local postbuckled states \cite{Plaut2015,Napoli2015,Pandey2014,Forterre2005}. Predicting and controlling buckling responses on more complex structures is an open and increasingly active problem in as diverse a range of applications as mechanical engineering \cite{Hu2015} to biological morphogenesis \cite{Nelson2016}.

Here, we demonstrate how the buckled states and buckling transitions of complex systems can be comprehensively surveyed and controlled. Our key methodological contribution is to combine a simple and versatile triangulated lattice model for modelling the shell morphologies with efficient high-dimensional free-energy minimisation and transition path finding algorithms, in order to develop a powerful computational methodology for exploring the buckling landscapes. We apply this approach to the problem of cylindrical shell buckling, where the extreme landscape complexity arises from a combination of subcritcality, multiplicity and snaking in the postbuckled states \cite{Knobloch2015}. Harnessing these tools we also explore the landscape biasing technique as an effective method to design and control buckling responses.

We begin by surveying the (meta) stable states - the free energy minima in the landscape. It has long been recognised that the cylindrical postbuckling states are strongly subcritical; coexisting with the unbuckled states in a loading interval which spans between the lower buckling load \cite{VonKarman1941}, and critical load (see \cite{Thompson2015} for a detailed review). This means that, as has been revealed historically, cylindrical shells are capable of failing at even 20\% of their critical load \cite{Seide1960}. Furthermore, the particular sensitivity to lateral loads \cite{Thompson2016}, lead to NASA's development of empirical predictions for the practical load bearing capacity of imperfect cylinders \cite{NASA1965}. Previously, buckled states have been elucidated by solving the von K\'{a}rm\'{a}n-Donell equations relating the stress to the radial displacement in an elastic cylindrical shell, but only by assuming the solutions exhibit axial periodicity (see for example \cite{Lord1997,Hunt2000}), reminiscent of the diamond pattern shown by Yoshimura to enable global, inextensible buckling \cite{Yoshimura1955}. Similarly, group theory has also enabled the study of high-symmetry solutions \cite{Wohlever1995}. However, a plethora of postbuckling solutions exist, discussed recently in the context of spatial localisation of the elastic deformation leading to snaking (pinning) in the solution space \cite{Knobloch2015}.

\begin{figure*}[!ht]
\centering
\includegraphics[width=\textwidth]{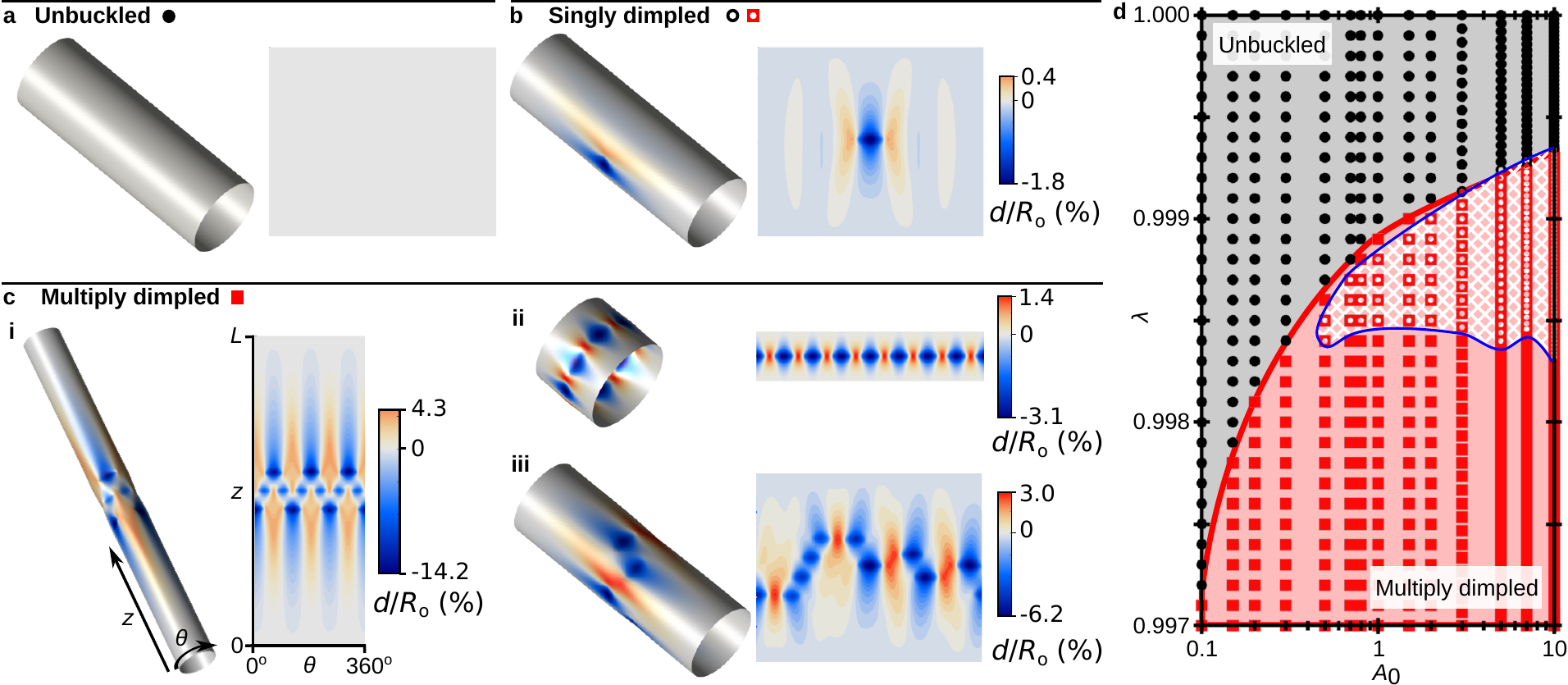}
\caption{Summary of the stable buckling morphology classes and the minimum energy states, surveyed across a range of aspect ratios $A_0$ and end shortening ratios $\lambda$. \textbf{a-c} Visualisations of representative minima, shown in 3D and as radial displacement field contour plots. The axial and angular coordinates $z$ and $\theta$ of the contour plots are shown in \textbf{c}(\textbf{i}), with the displacement $d$ expressed as a fraction of $R_0$. \textbf{d} Phase diagram indicating the global free energy minimum across a range of $A_0$ and $\lambda$. The control ratio, $k^{\rm{stretch}}R_0^2/k^{\rm{bend}}$ is fixed throughout at $2.5 \times 10^{5}$. The global minimum is either unbuckled (grey region, simulation data shown as black circles), or multiply dimpled (pink region, simulation data shown as red squares). The singly dimpled state is never the global minimum, but the existence region is shown outlined in blue with unfilled data points.}
\label{fig:1phasediag}
\end{figure*}

The energy landscape approach is then used to survey how the minima transform into each other by the lowest energy routes. We connect all postbuckling morphologies via such transition pathways, and so explore the entire stability landscape of cylindrical buckling. In this we reveal a diverse variation in the landscape properties, ranging from very simple funnel-shaped landscapes at low aspect ratios and end shortenings, to broad and highly complex glassy landscapes at long aspect ratios.

Previously, only the first transition capable of buckling the unbuckled cylinder has been investigated \cite{Horak2006, Kreilos2017}. This has received significant interest as capturing the minimum energy pathway (MEP) enables the minimum energy barrier to be obtained, which provides an absolute lower bound to the energy required for a compressed cylinder to buckle. An explicit link has therefore been made between the ease of single dimple formation and the sensitivity of loaded cylinders to lateral loading \cite{Horak2006}. As important for structural applications, it has been suggested that these theoretical minimum energy barriers can be accessed experimentally via a local probing technique for cylindrical  \cite{Virot2017,Gerasimidis2018,Thompson2016, Thompson2017} and spherical shells \cite{Hutchinson2017,Marthelot2017}. However, it is only via comparison to the minimum energy pathways obtained here that we are able to verify this. 

Finally, we introduce a new method to begin to exert control over the buckling landscape - landscape biasing. In this, we are able to stabilise or destabilise targeted features in the landscape, such as transition states and minima. This is achieved by making local modifications to the elastic spring constants in the triangular lattice model to simulate thickness modifications. Thus, the knowledge of the energy landscape proves highly complimentary to experimental processes aimed at exerting postbuckling control \cite{Hu2015a,Hu2016,Kuang2019}. We demonstrate the principal of landscape biasing by first showing how biasing against the unbuckled-single dimple transition state produces a 20\% increase in buckling resistance of the unbuckled cylinder for a 1\% increase in mass. We then show how biasing for a multiply dimpled state simplifies the local landscape, tripling the targeted state stability at 0\% mass change.

\section{Results and Discussions}

\subsection{Free energy minima}

The triangular lattice model, detailed in Methods, discretises the shell into a triangulated mesh of extensional and angular elastic springs. Respectively, these allow for the decomposition of the total free energy into a sum of stretching and bending terms. To begin with, the stretching and bending spring constants, $k^{\rm{stretch}}$ and $k^{\rm{bend}}$, are uniform throughout the shells. Each shell is generated with a well-defined aspect ratio $A_0=L_0/(2R_0)$, where $L_0$ and $R_0$ are the length and radius of the cylinder when all springs assume their equilibrium configurations. When axially compressed, the shortening ratio is defined $\lambda = L/L_0$, where $L$ is the length of the compressed cylinder. The top and bottom edges of the cylinder are simply supported: the coordinates of the mesh are fixed, but the planes attached to the ends can bend freely. We also choose $k^{\rm{stretch}}$, $k^{\rm{bend}}$, and $R_0$ to maintain a constant dimensionless elastic control ratio $k^{\rm{stretch}}R^2_0/k^{\rm{bend}}$=$2.5\times10^5$, and free energies $E$ reported throughout are nondimensionalized such that the reduced free energy $E_{\rm{r}}=E/k^{\rm{bend}}$. The elastic control ratio is chosen to be representative of a physical system, corresponding to an aluminium drinks can. The choice of elastic control ratio and nondimensionalization are discussed further in Methods.

The construction of the free energy landscape begins by surveying the free energy minima. To access the many different buckled states for each fixed aspect ratio $A_0$ and shortening ratio $\lambda$, a basin hopping step is employed prior to energy minimisation \cite{Wales2003}, detailed in Methods. Three characteristic cylinder morphologies are observed: unbuckled, singly dimpled, and multiply dimpled, visualised in 3D and as radial displacement fields $d$ in Fig. \ref{fig:1phasediag}a-c respectively.

The multiply dimpled states in Fig. \ref{fig:1phasediag}c form the largest set of minima, within which is contained the often-studied morphologies of high rotational symmetry - the Yoshimura-like diamond dimpling pattern, an example of which is shown in Fig. \ref{fig:1phasediag}c(ii) \cite{Yoshimura1955}. However the largest multiply dimpled subset is the irregularly dimpled morphologies, a characteristic example of which is shown in Fig. \ref{fig:1phasediag}c(iii). A random perturbation applied to these cylindrical shells is therefore most likely to result in an irregularly dimpled state, showing that cylindrical shell buckling responses are inherently hard to predict.

The phase diagram in Fig. \ref{fig:1phasediag}d summarises the minimum survey. At high $\lambda$ in the phase diagram, the global free energy configuration is the unbuckled state, indicated by black circles. Upon decreasing $\lambda$ the multiply dimpled states become the global minima, indicated by red squares. The solid red line indicates the point at which the buckled and multiply dimpled states are isoenergetic, at which the shortening ratio therefore produces an axial load equal to the Maxwell load (a detailed discussion regarding the loading limits is given in \cite{Thompson2015} for example). Across all tested scenarios, the most stable multiply dimpled states are those exhibiting a high degree of rotational symmetry, most commonly those with Yoshimura-like diamond patterns. However, we also find examples where more exotic high-symmetry multiply dimpled states form the global free energy minima, such as the example shown in Fig. \ref{fig:1phasediag}c(i) at $A_0=10, \lambda=0.999$.

The singly dimpled state, shown in Fig. \ref{fig:1phasediag}b is of significant interest due to it's frequent role in the first buckling transition (which we consider further in the proceeding section), and also it's characteristic role of being the unit excitation in the postbuckling landscape. However, across a broad range of aspect ratios and elastic control ratios, detailed further in Supplementary Note 1, we observed that the single dimple is never the global free energy minimum. When it is energetically unfavourable to form a dimple, the unbuckled state is lower in energy; when it is energetically favourable to form a dimple, the energy is always lowered further by subsequent dimpling. The single dimple is therefore only metastable. This metastability region is outlined in blue in Fig. \ref{fig:1phasediag}d. The non-monotonic form of the low-$\lambda$ boundary arises from the complex deformation profile surrounding the dimple. At high aspect ratios, this profile extends around the circumference of the cylinder, such that self-interaction effects contribute to the dimple stability (detailed further in Supplementary Note 1).

\subsection{Buckling transitions}
\begin{figure}[!ht]
\centering
\includegraphics[width=0.5\textwidth]{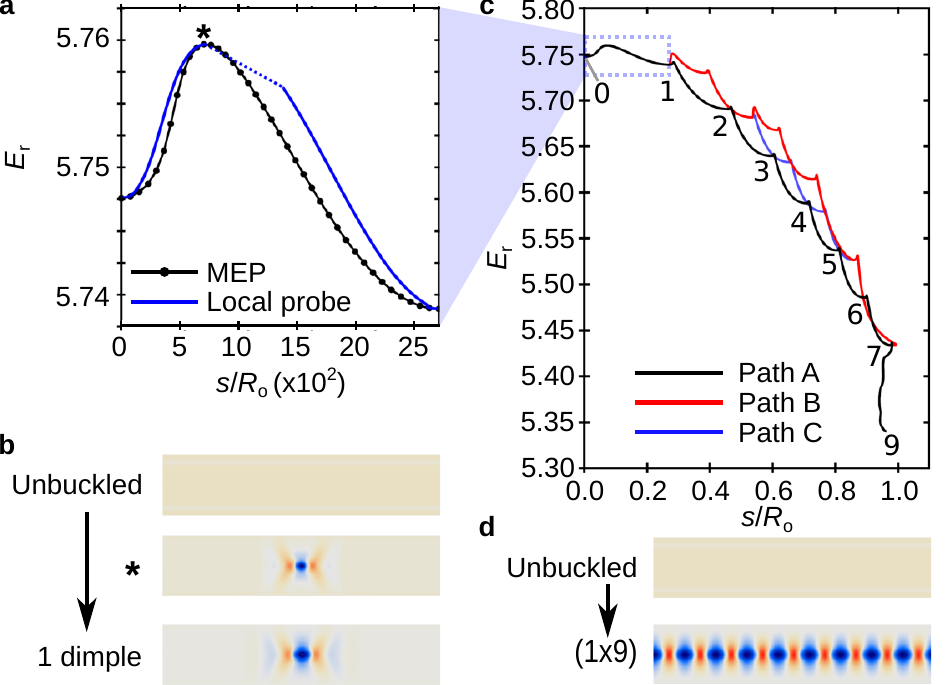}
\caption{Comparison of the minimum energy pathway with the local probe technique, and examples of multi-step pathways through the buckling landscape. \textbf{a} Reduced free energy profile along the MEP (black) and local probe pathway (blue) for the unbuckled - single dimple transition on a cylinder of $A_{0}=0.8$, $\lambda=0.9986$. The end points and transition state (*) are illustrated in \textbf{b}. The path length $s$ describes the normalised distance of a point along the profile from the unbuckled state shown in Eq. \eqref{eq:distance}. \textbf{c} Three example transition pathways connecting the unbuckled state and 1-row by 9-dimples (1$\times$9) state illustrated in \textbf{d}. All pathways are shown to begin with the unbuckled - single dimple transition, which is magnified in \textbf{a}, and the number of dimples are labelled at each minimum in Path A. }
\label{fig:2multipath}
\end{figure}

In order to describe the minimum energy mechanisms by which the cylindrical buckling morphologies interconvert, we must obtain the minimum energy pathways (MEP). Between any two states in the free energy landscape, the MEP is defined as a path in which the gradient of the free energy is parallel to the path tangent vector. The MEP will also pass through at least one saddle point in the landscape, a local energy maximum along the pathway. The buckling morphology at this point is known as the transition state. Several methods exist for finding the MEP and transition states, see for example refs. \cite{Sheppard2008, Ren2013, Trygubenko2004, Henkelman2000, Kusumaatmaja2015}. The string methods we use here are detailed in Methods.

Computationally, the only transition which has been followed previously is the simplest unbuckled-singly dimpled pathway, where the dimple is centrally located on the cylinder \cite{Horak2006, Kreilos2017}. Meanwhile, local probing of cylindrical shells has been suggested as an experimental technique which may allow the true dimpling transition state to be accessed \cite{Virot2017,Gerasimidis2018,Thompson2016, Thompson2017}. In Fig. \ref{fig:2multipath}a, we compare the reduced-energy profiles $E_{\rm{r}}(s)$  of the MEP (black series) with the pathway generated by simulating the local probe technique (blue line) for an example cylinder with $A_0=0.8, \lambda=0.9986$. Local probe simulation methodologies are detailed in Methods. In order to usefully compare the paths, the path distance coordinate \textit{s} is the Euclidean distance between the triangulated mesh of a point along the pathway, with that of the initial (unbuckled) state
\begin{equation}
s=\sum_{i=1}^{N_{\rm{nodes}}} \vert \underline{\mathbf{a}_i} - \underline{\mathbf{a}_i^o} \vert
\label{eq:distance}
\end{equation}
where $\underline{\mathbf{a}_i}$ and $\underline{\mathbf{a}_i^o}$ are the position vectors of node $i$ in the buckled and unbuckled mesh respectively.

On comparison, we observe that the local probe technique does meet the MEP at the transition state (labelled '*' and shown in shown in Fig. \ref{fig:2multipath}b), but does not access the minimum energy pathway generally. At the point of crossing the barrier, the locally-probed system snaps to a dimpled-like configuration: a small probe displacement resulting in a large change to the surrounding morphology, and a concomitant jump in $E_{\rm{r}}$ and $s$. 

This comparison shows that the local probe technique is capable of measuring the minimum energy barrier to the first dimpling transition. This is consistently shown across all the test cases summarised in Supplementary Note 2. Previous studies were unable to prove that the local probe technique could access the minimum energy barrier, as it was assumed that the true MEP was not too curved \cite{Thompson2016}: namely the direction of motion along the transition always has a component in the direction of the applied force (i.e the path never curves against the applied force).

However, the methodology presented here allows for the pathway between any two states to be investigated, not only the 0-1 transition. We therefore extend the first pathway found in Fig. \ref{fig:2multipath}a to find complete pathways from the unbuckled state to the multiply-dimpled global minimum. Examples are shown in Fig. \ref{fig:2multipath}c, with Fig. \ref{fig:2multipath}d showing the pathway endpoints: the unbuckled state, and the (1$\times$9) global minimum. Two key observations are made: multiple competing pathways exist between the end points, and each pathway is complex, featuring many intervening minima. Out of the large number of possible pathways, three examples are highlighted in Fig. \ref{fig:2multipath}c, labelled  A, B, and C. Movies showing the conformational changes along each pathway are shown in Supplementary Movies 1, 2, and 3 respectively. Path A is distinguished from other paths: out of the set of barriers along path A, the maximum energy barrier is the smallest out of all possible pathways. In Path A, eight separate dimpling transitions occur. In the first seven, a single dimpling event occurs to build a train of dimples. The final transition sees two dimples forming simultaneously to complete the ring of nine dimples. In this final transition, the path distance decreases as all dimples become shallower on formation of the final two. However, the system is capable of undergoing dimpling transitions not linked to the growing dimple train, leading to example alternative pathways B and C.

\subsection{Energy landscapes}
\begin{figure*}[!ht]
\centering
\includegraphics[width=\textwidth]{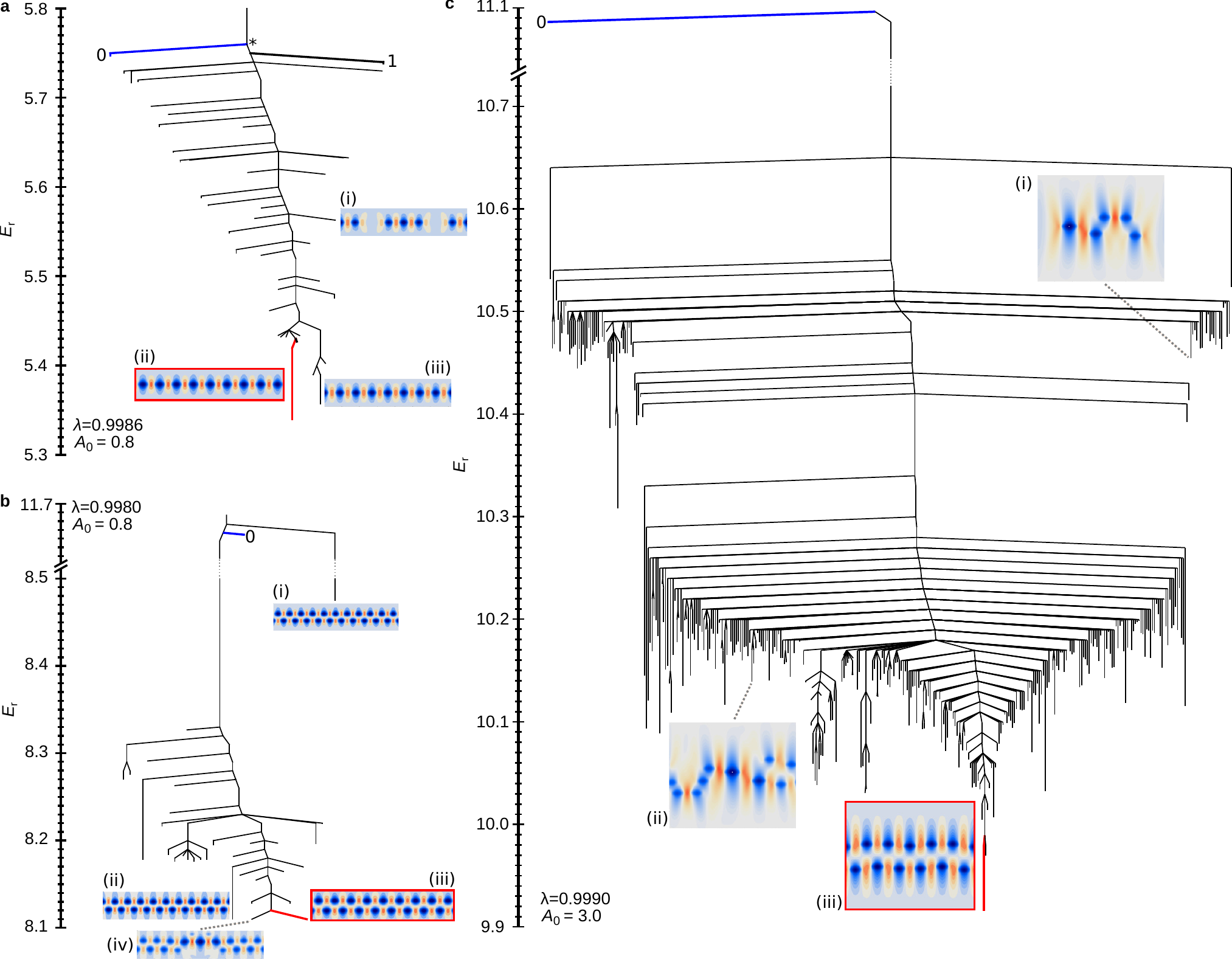}
\caption{Disconnectivity graphs showing the minimum energetic barrier between any pair of states. The unbuckled and global minimum branches are coloured in blue (labelled '0') and red respectively. Representative minima radial displacement plots are also shown, with the global minimum outlined in red. \textbf{a} $A_0$ = 0.8, $\lambda$ = 0.9986, the single dimple branch is labelled '1', the 0-1 transition state labelled '*'. \textbf{b} $A_0$ = 0.8, $\lambda$ = 0.9980, no additional minima are present in the vertical axis break. \textbf{c} $A_0$ = 3.0, $\lambda$ = 0.999, a small number of multiply dimpled states not pertinent to the discussion are present in the vertical break.}
\label{fig:3dis}
\end{figure*}

By connecting any pair of minima with an MEP, we may thus explore the complete energy landscape for any fixed $A_0$ and $\lambda$. Here we examine the extent of the landscape complexity as a function of $A_0$ and $\lambda$ (varying the elastic constants is presented in Supplementary Note 3). As will be shown, cylindrical shells exhibit a diverse range of landscape types. We will first compare the energy landscape of a lightly compressed short cylinder where the single dimple is stable ($A_0=0.8$,$\lambda=0.9986$), with a heavily compressed short cylinder where the single dimple is unstable ($A_0=0.8$,$\lambda=0.9980$). We then compare the short, lightly compressed cylinder, with a long, lightly compressed cylinder ($A_0=3.0$,$\lambda=0.9990$), where the single dimple is stable in both cases. 

As the network of minima connected by MEPs is in general highly complex, it is instructive to consider simplified network representations. In Fig. \ref{fig:3dis}, the free energy landscapes are visualised as disconnectivity graphs (for a comprehensive discussion of the disconnectivity graph representation of energy landscapes, we refer the reader to refs. \cite{Wales2003, Wales1998}). In this, the network of minima and pathways is reduced to a spanning tree showing only the energy of the minima (the end points of each branch) and the lowest energy barrier connecting any two minima, read by tracing the path between two branches and finding the highest energy point. For example, in Fig. \ref{fig:3dis}a, unbuckled state and singly dimpled state are labelled '0' and '1' respectively. On tracing between the two branches, the highest energy point along the path, labelled '*' marks the largest transition state energy. In this case, this is the 0-1 transition state shown in Fig. \ref{fig:2multipath}b. However, as the 1D disconnectivity graph does not show which states are directly connected, in general the highest energy point between two states is simply the largest energy encountered in the possible multi-step transition pathway.

In Fig. \ref{fig:3dis}a, the disconnectivity graph is presented for $A_0$ = 0.8, $\lambda$ = 0.9986, and represents the full energy landscape which was partially described in Fig. \ref{fig:2multipath}. Under these sub-critical conditions, the unbuckled, singly dimpled, and multiply dimpled states coexist. However, the buckling landscape is remarkably simple: qualitatively, the states are (approximately) uniformly distributed across the stable energy range. To quantify this and subsequent observations, we partition the minimum-energy range into 100 bins of equal width and total the number of minima within each bin; this histogram is shown in Supplementary Note 4. We then calculate the variance in bin populations as a measure of the distribution uniformity. Here, the small variance in the bin frequency, 0.38, describes a relatively uniform distribution of minima across the energy range. The uniformity of the landscape is further reflected in the range of energy barriers - almost all have similar minimum energy barriers, of energy $\mathcal{O}(10^{-3})$. The distribution of the barriers is also shown in Supplementary Note 4.

The example minimum (i) is a characteristic state of the system, featuring clusters of dimples closely aligned around the central circumference. The (1$\times$9) global energy minimum (highlighted in red) exists in a deep well, with the minimum energy barrier greater than the first transition by a factor of 7. Thus, if an unbuckled state is subject to perturbations with sufficient energy to overcome the first dimpling transition, although other states may be sampled along the way, the tendency is to quickly become trapped in the global energy minimum. The notable exception to this picture however is that a second deep branch also exists at the base of the disconnectivity graph. This represents a competing set of deep states which are likely to split the population between the lowest minimum (1$\times$9), and second-lowest minimum (1$\times$8), labelled (iii) in Fig. \ref{fig:3dis}a.

Upon decreasing  $\lambda$ to 0.9980, although the system is still subcritical, the singly dimpled state looses stability. The disconnectivity graph for this landscape is shown in Fig. \ref{fig:3dis}b. Here, the landscape is markedly different to the less-compressed case shown in Fig. \ref{fig:3dis}a: although the number of minima is $\mathcal{O}(10)$ in both cases, at $\lambda$ = 0.9980 the majority of states are concentrated at the lower stable energy range, indicated by the greater variance in bin population, 2.09, detailed further in Supplementary Note 4. Additionally, the range of energy barriers is large, varying from $10^{-3}$ to $10^{1}$, with many states featuring high energetic barriers. This latter point is most pronounced when considering the (2$\times$11) multiply dimpled state, labelled (i), which has an energy barrier 1000$\times$ greater than the minimum energy barrier from the unbuckled state. A further contrast in this disconnectivity graph is that the global minimum (2$\times$9) does not have a large energy barrier compared to other transitions. Thus, random perturbations made to the unbuckled state may result in the system becoming trapped in several states different from the global minimum. Two highlighted examples of these which are close in energy to the global minima are the (2$\times$10) system, labelled (iii), and a defective (2$\times$9) system with two adjacent dimple vacancies, labelled (iv).

Finally, we return to a subcritical shortening ratio where the unbuckled, singly dimpled, and multiply dimpled states coexist, but now extend the aspect ratio: $A_0$ = 3.0, $\lambda$ = 0.999. The disconnectivity graph for this system is shown in Fig. \ref{fig:3dis}c. Three prominent features of this landscape offer significant contrast to the short aspect-ratio landscapes: the number of minima has increased by a factor of 100 compared to the $A_0$ = 0.8 systems, the minimum distribution is highly non-uniform  - the bin population variance is 93, and the landscape becomes rough over a range of energy scales.

Expanding on these observations, the increase in the number of minima is due to two effects. Firstly, at large aspect ratios, all minima observed are no longer characterised uniquely by a single well-defined energy and morphology, but exist as clusters in which the intra-cluster energy variability is approximately $\Delta E_{\rm{r}} < \times 10^{-3}$. Thus, on the finest scale, the stability landscape is rough and glass-like. In the stability landscape shown in Fig. \ref{fig:3dis}c, we have clustered minima which share the same number of dimples with interconversion barriers $< 10^{-3}$, reducing the number of minima shown by a factor of 10. The second effect is due to dimple confinement introduced by the fixed ends. At $A_0$ = 0.8, the fixed ends tightly constrain the dimples to lie within either one or two rows, due to the characteristic dimple size being similar to $L_0$. At the longer aspect ratio of $A_0$ = 3.0, the constraining strength of the fixed ends is diminished, yielding a larger number of possibilities of dimple arrangements. 

\begin{figure*}[!ht]
\centering
\includegraphics[width=\textwidth]{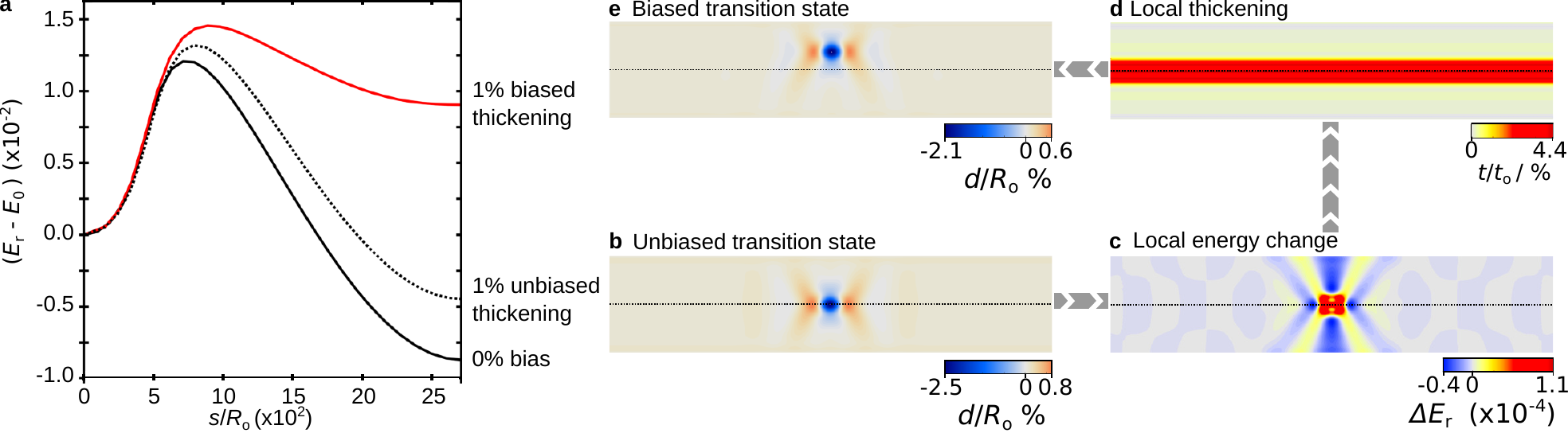}
\caption{The landscape biasing workflow and the effect when biasing against the unbuckled-single dimple transition state. \textbf{a} Unbuckled to single dimple transition energy profiles for three local thickening schemes: $A_0$=0.8, $\lambda$=0.9986. \textbf{b-e} Illustrative workflow for the landscape biasing procedure, a black dotted line indicates the centre of the cylinder. \textbf{b} Radial deformation field of the unbiased transition state. \textbf{c} Local elastic potential energy change of the transition state relative to the unbuckled cylinder. \textbf{d} Local thickening profile of the 1\% biased cylinder. \textbf{e} Unbuckled to single dimple transition state of the  1\% biased cylinder.}
\label{fig:4nudge}
\end{figure*}

The large phase space for dimple arrangements within certain energy ranges enables numerous minima to exhibit similar energies and similar barriers. This is most pronounced in the range $10.1 < E_{\rm{r}} < 10.3$, dominated by irregular systems with between 7 and 11 dimples. A representative example is shown, labelled (ii). In this region, the number of dimples is large enough to produce a significant number of variations in arrangement, yet not so large that packing constraints become dominant. On average, the inter-cluster energy barrier is $\mathcal{O}(10^{-2})$. A similar glassy region exists at larger energies, where irregularly dimpled systems feature between 3 and 6 dimples. A representative example here is shown, labelled (i). Thus, the stability landscape becomes rough on two energy scales: (1) $\Delta E_{\rm{r}} \approx \times 10^{-3}$ associated with intra-cluster variability, and (2) $\Delta E_{\rm{r}} \approx \times 10^{-2}$ associated with inter-cluster variability in the absence of packing constraints (when comparing clusters of similar numbers of dimples). The distributions of energy barriers associated with this roughness are shown in the Supplementary Note 4.  

For larger dimple numbers than 11, efficient packing on the cylinder is required, leading to a severe reduction in the phase space of dimple arrangements. Thus, in the vicinity of the global minimum, the (2$\times$6) regularly dimpled state highlighted in red, the local landscape becomes significantly less glassy. Nonetheless, the overall landscape roughness coupled with a large number of deep states means that a perturbed unbuckled cylinder may buckle to any number of states, explaining the difficulty in designing cylindrical postbuckling states.

\subsection{Controlling the landscape}

\begin{figure*}[!ht]
\centering
\includegraphics[width=\textwidth]{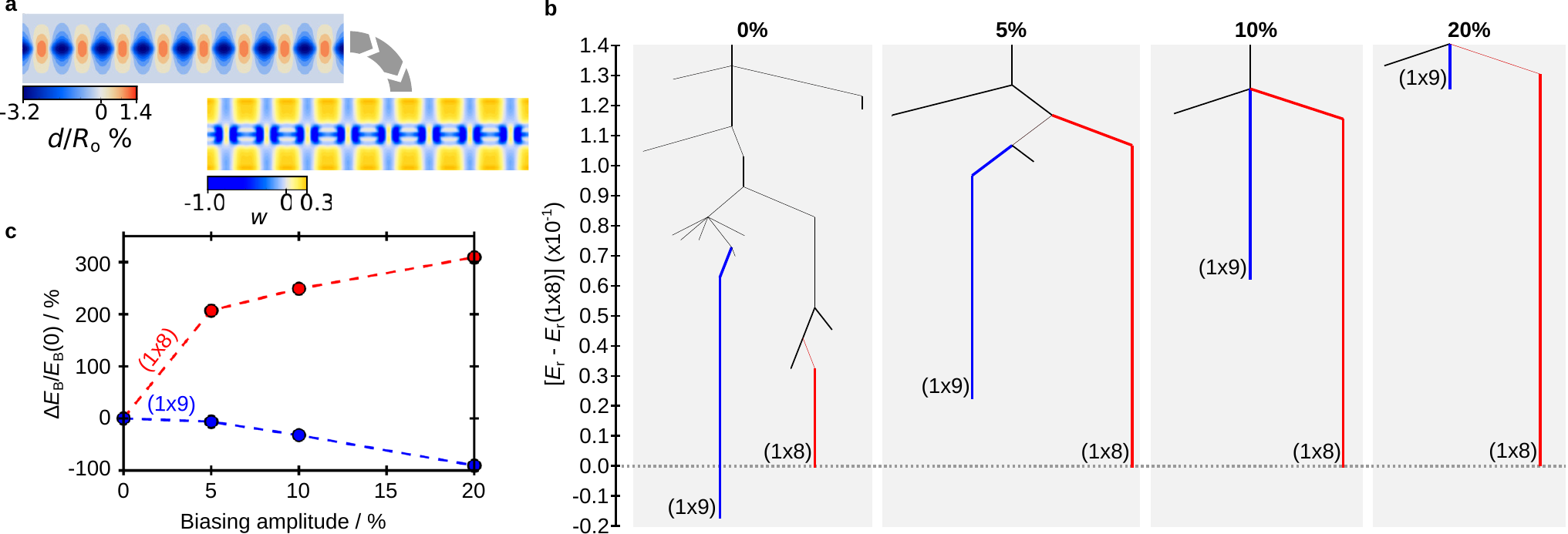}
\caption{Changes in the local landscape upon biasing for the (1$\times$8) state. \textbf{a} Workflow showing how the radial deformation field of the (1$\times$8) state at $A_0$=0.8, $\lambda$=0.9986 is transformed into the thickness weighting field for landscape biasing. \textbf{b} Evolution of the bottom of the landscape as the biasing amplitude increases, all energies shown relative to the (1$\times$8) state. The (1$\times$8) well is highlighted in red and the (1$\times$9) well is highlighted in blue. \textbf{c} Evolution of the change in minimum energy barrier $\Delta E_{\rm{B}}$ out of the (1$\times$8) state (red), and (1$\times$9) state (blue) upon increase in biasing amplitude. $\Delta E_{\rm{B}}$ is shown relative to the unbiased barrier, $E_{\rm{B}}(0)$, for the (1$\times$8) state and (1$\times$9) state respectively. The dotted lines are shown as guides for the eye.}
\label{fig:51x8}
\end{figure*}

Despite the complexity of the buckling landscapes, we now demonstrate how to control the stability of target features, by introducing a process we term \textit{landscape biasing}. This enables us to design buckling responses by locally thickening or thinning the cylinder, complimentary to experimental realisation; see for example \cite{Hu2015a,Hu2016,Kuang2019}. We demonstrate two examples of landscape biasing, by first biasing \textit{against} a target transition state, and then biasing \textit{for} a target minimum. The examples shown here significantly increase the stability of the target structures to lateral perturbations. These biased structures are therefore highly suited to scenarios where sudden morphological changes would be detrimental to device performance, a key example being aeronautical applications \cite{NASA1965}. For these examples, we apply this method to $A_0$=0.8, $\lambda$=0.9986 system, for which the buckling landscape is shown in Fig. \ref{fig:3dis}a.  

To begin with, it is observed that the minimum energy barrier from the unbuckled state to the singly dimpled state is small compared to both the overall landscape energy range, and other deep states, generating the extreme imperfection sensitivity of cylinders to sub-critical buckling transitions. The energy profile for this transition, shown originally in Fig. \ref{fig:2multipath}a, is re-plotted in Fig. \ref{fig:4nudge}a (solid black line), in which the reduced energy is referenced to the energy of the unbuckled state, $E_o$. We aim to increase the energy barrier of this transition, in order to make the unbuckled cylinder more robust against lateral perturbations, by biasing the landscape against the transition state.

The landscape biasing workflow is shown in Fig \ref{fig:4nudge}b-d, and detailed further in Supplementary Note 5. Firstly, as shown in Fig. \ref{fig:4nudge}b, we obtain the radial deformation field for the unbiased transition state (as well as that of the unbuckled state). Secondly, we compute the fractional change in local elastic potential energy $E_{\rm{f}}$ when transforming from the unbuckled to the transition state. It is observed that the stored elastic potential energy is highly localised about the centre of the dimple deformation. We then reason that in order to increase the energy of this transition state (and hence the barrier to the transition), we must modify the cylinder to energetically penalise this localisation of the potential energy, effectively biasing the landscape against the transition state. A choice exists in how to perform this modification, but for this example we choose to simulate a local thickening of the shell by modifying $k^{\rm{stretch}}$ ($\propto t$) and $k^{\rm{bend}}$ ($\propto t^3$), facilitating experimental realisation. A more sophisticated yet complex treatment would alter $k^{\rm{stretch}}$ and $k^{\rm{bend}}$ independently, according to the separate local stretching and bending energies respectively. A comparison of alternative geometric methods to modify cylindrical shell buckling are presented in \cite{Hu2015a}. In the local thickening treatment, detailed in Supplementary Note 5, we weight the thickening according to the local energy change. Due to the symmetry breaking of the transition, in order to suppress dimple formation anywhere around the circumference of the cylinder, at each $z$ we average the thickening profile over all $\theta$. Finally, the thickening profile is rescaled in order to achieve a prescribed total mass increase, which is set as 1\% for the results presented in Fig. \ref{fig:4nudge}. The final thickening profile is shown in Fig. \ref{fig:4nudge}d, which sees the a thickness increase localised around the centre of the cylinder.

On attempting to dimple this biased cylinder, the transition state is now forced off-centre, shown in Fig. \ref{fig:4nudge}e. The energy profile for this transition is shown as the solid red line in Fig. \ref{fig:4nudge}, showing that for a 1\% increase in mass, a 20\% increase in buckling resistance is achieved. This improvement is over twice that of a uniformly thickened cylinder, 9\%, with the same mass increase, the transition profile for which is shown as the dotted black line. This landscape biasing against the transition is the antithesis to modal nudging \cite{Cox2018}, the recently formalised technique for slender structures in which minimal structural modifications are made in order to select a specific failure mode. 

The second way to design the bucking landscape is to bias \textit{for} a target structure. We observe the landscape shown in Fig. \ref{fig:3dis}a to exhibit a deep global minimum (1$\times$9) and the shallower (1$\times$8) state. Here, we choose to stabilise the (1$\times$8) state through minimum-targeted landscape biasing. It will be shown how a target minimum can be significantly stabilised, thus realising a postbuckled state which is highly resistant to lateral perturbation. Furthermore, this example will show that through biasing we can select which high-symmetry morphology forms the global minimum. 

In Fig. \ref{fig:51x8}a, we show the radial displacement field of the (1$\times$8) state. As before, we evaluate the local stored elastic potential energy, then weight the local elastic constants to exact a local thickening, detailed further in Supplementary Note 5. As the (1$\times$8) state is to be stabilised, in regions of high stored elastic energy we locally thin the structure to reduce the energetic cost of the specific buckling mode. We also weight the thickening so that there is no overall mass change, and prescribe a biasing amplitude - the maximum percentage change in thickness allowed. To obtain the local thickness change, we therefore scale the weighting field $w$ shown in fig. \ref{fig:51x8}a by the biasing amplitude.

By systematically increasing the biasing amplitude from 0\% to 20\%, we observe how the buckling landscape changes at the bottom of the funnel, shown in Fig. \ref{fig:51x8}b. At 0\% bias, we show a magnification of the low-energy portion of the disconnectivity graph shown in Fig. \ref{fig:3dis}a, featuring the two deep wells decorated with multiple stable minima. The wells corresponding to the (1$\times$8) state and (1$\times$9) state are shown highlighted in red and blue respectively. In Fig. \ref{fig:51x8}c, the percentage change in the (1$\times$8) and (1$\times$9) barriers are shown relative to their respective barriers at 0\% bias.

On application of a 5\% bias, the landscape changes significantly relative to the unbiased case: the landscape is simplified as the biasing destabilises many minima, the (1$\times$9) state increases in energy, and the targeted (1$\times$8) state decreases in energy to such an extent that it becomes the global minimum. Furthermore, the landscape simplification and (1$\times$8) state stabilisation effects act cooperatively to increase the barrier out of the target (1$\times$8) state by 207\% relative to the unbiased (0\%) landscape. At 10\% bias, these effects are further magnified. At 20\% bias, there is no further change in the lower landscape structure, but the stabilisation of the (1$\times$8) state and destabilisation of the (1$\times$9) state continues. This leads to an ultimate barrier increase 302\% for the (1$\times$8) state, and barrier decrease of 91\% for the (1$\times$9) state.

\section{Conclusions and Outlook}
In this work, a triangular lattice model is used to evaluate the free energy of postbuckled states of elastic thin shells. This is implemented in efficient energy-minimisation and path finding algorithms in order to fully describe the buckling landscapes. Here, we have demonstrated this for the complex problem of buckling of fixed-end cylindrical shells, subject to axial compressive strains. To begin with, we surveyed the free energy minima, observing unbuckled, singly dimpled, and multiply dimpled states whose stabilities were evaluated for different aspect ratios and compressive strains. We then systematically used the string method to connect pairs of minima within the same cylindrical system in order to find the minimum energy pathways and transition states between these states. This enabled a global description of the buckling landscape: in which a simple funnel-shaped landscape became complex and glassy when increasing the aspect ratio, or featured many deep states when increasing the compressive strain. We then finally introduced the landscape biasing method to control the stability of targeted features of the landscape, in order to design structures with improved resistance to lateral forces. 

Overall, by being able to both survey the free energy landscape and design specific transition modes through landscape biasing, we may now design dynamic buckling responses for diverse applications, ranging from energy harvesting devices to complex morphable materials.

One important consideration we highlight for future work is that of the role of imperfections in buckling responses, a significant concern in real-world applications. The ability to generalise our model to consider shapes other than the perfect cylinder, as well as including diverse elastic modulations and boundary conditions, lead us to emphasise the applicability of this model to studying the impact of a large range of different geometric or elastic imperfections on the buckling landscape.

\section{Methods}

\subsection{Discretisation and free energy}
\begin{figure}[!ht]
\centering
\includegraphics[width=0.5\textwidth]{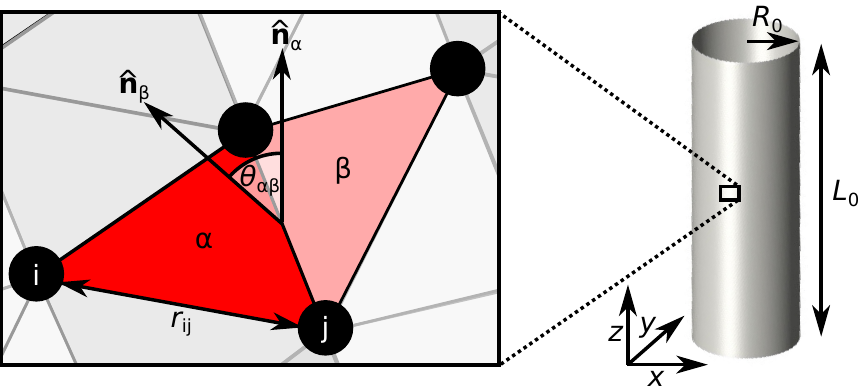}
\caption{The thin shell discretisation scheme. The nodes are indicated with black circles, in which nodes i and j are separated by a distance $r_{\rm{ij}}$. The planes are indicated with coloured triangles, in which the dihedral angle between planes $\upalpha$ and $\upbeta$, $\theta_{\rm{\upalpha \upbeta}}$, is shown as the angle between the respective normal vectors $\hat{\textbf{n}}_{\rm{\upalpha}}$ and $\hat{\textbf{n}}_{\rm{\upbeta}}$. }
\label{fig:5methods}
\end{figure}

To evaluate the free energy of an arbitrary thin shell (or composite of thin shells), we discretise the surface into a triangulated mesh of nodes, defining a set of neighbouring nodes and a set of neighbouring planes, in a manner based on \cite{Seung1988} although other similar methods have also been reported, for example \cite{Liu2017}. The local form of this discretisation is shown in Fig. \ref{fig:5methods}. In this, neighbouring nodes i and j are connected by an extensional spring of equilibrium bond length $r^{0}_{\rm{ij}}$ and elastic constant $k^{\rm{stretch}}_{\rm{ij}}$. Neighbouring planes $\upalpha$ and $\upbeta$ are connected by an angular spring of equilibrium angle $\theta^{0}_{\rm{\upalpha \upbeta}}$ and elastic constant $k^{\rm{bend}}_{\upalpha \upbeta}$. In general, as in our triangulation scheme, $r^{0}_{\rm{ij}}$ and $\theta^{0}_{\rm{\upalpha \upbeta}}$ are non-uniform across the lattice. The discretisation of the shell into a set of extensional and angular springs allows the total free energy to be decomposed into a sum of stretching and bending energies such that generally,
\begin{align}
E&=\sum_{\rm{ij}}k^{\rm{stretch}}_{\rm{ij}}\left( r_{\rm{ij}}-r^{0}_{\rm{ij}} \right)^2 \nonumber \\ 
&+ \sum_{\upalpha \upbeta} k^{\rm{bend}}_{\upalpha \upbeta} \left( 1- \cos \left( \theta_{\rm{\upalpha \upbeta}} - \theta^{0}_{\rm{\upalpha \upbeta}} \right) \right),
\end{align}
where $r_{\rm{ij}}$ is the separation distance between nodes i and j; $\theta_{\rm{\upalpha \upbeta}}$ is the dihedral angle between planes $\upalpha$ and $\upbeta$, defined as the angle between the respective normal vectors $\hat{\textbf{n}}_{\rm{\upalpha}}$ and $\hat{\textbf{n}}_{\rm{\upbeta}}$.

Throughout this work, we report the nondimensionalised free energy $E_{\rm{r}} = E / k^{\rm{bend}}_{\rm{ref}}$, where $k^{\rm{bend}}_{\rm{ref}}$ is a reference dihedral elastic constant. For cylinders of uniform elasticity, we define $ k^{\rm{bend}}_{\rm{ij}} = k^{\rm{bend}}_{\rm{ref}}$. Furthermore, the bond lengths are nondimensioanlised by expressing $r_{\rm{ij}}$ and $r^{0}_{\rm{ij}}$ relative to a reference length scale $R_0$, which we choose to be the cylinder radius.

The single parameter defining the cylinder's elastic behaviour then becomes the control ratio $ k^{\rm{stretch}}_{\rm{ref}} R_0^2/ k^{\rm{bend}}$ which unless otherwise stated we fix at $2.5\times10^{5}$. Through comparison with continuum elastic theory \cite{Seung1988}, in terms of Young's modulus $Y$, plate thickness $t$, and Poisson ratio $\nu$ we have $k^{\rm{stretch}}=\frac{\sqrt{3}}{4}Yt$ and $k^{\rm{bend}}=\frac{2}{\sqrt{3}}\frac{Yt^3}{12(1-\nu^2)}$. Hence, the control ratio is given by $\frac{9(1-\nu^2)}{2}\left(\frac{R_0}{t} \right)^2$. To demonstrate the physical significance of our prescribed control ratio of $2.5\times10^{5}$, if we choose a Poisson ratio appropriate for aluminium, $\nu=0.3$, the resulting ratio $R_0/t$=247 is similar to that of aluminium drinks cans ($R_0/t$ $\approx$ 300).

The cylinder radius $R_0$ is fixed throughout, such that to change the uncompressed aspect ratio $A_0$, only the length $L_0$ is varied. In order to accurately calculate the free energy while balancing computational cost, the number of nodes must be sufficient to capture the deformation profiles of single dimples, the length scale of which depends on $A_0$ and the control ratio. For the cylinders studied here, $\approx 10^4$ nodes per cylinder are required (an illustrative resolution test is shown in Supplementary Note 6). Our triangulated lattice model is also validated against ABAQUS/Explicit commercial software \cite{ABAQUS2001}, shown in Supplementary Note 7.

\subsection{Minimisation and path finding}
The L-BFGS algorithm \cite{Nocedal1980, Liu1989} is employed to efficiently minimise the free energy with respect to the large number of degrees of freedom ($\mathcal{O}(10^4-10^5)$). For this, the total free energy is required as well as the derivatives of $E$ with respect to each degree of freedom (the $x$, $y$ and $z$ coordinates of each node). By setting selected derivatives to zero prior to minimisation, we can constrain specific node positions. Here, we fix the $x$ and $y$ coordinates of the nodes which cap each end of the cylinder to the uncompressed configuration, forbidding deformation  or relative rotation of the ends. By choosing the $z$ coordinates at which to fix these nodes, we can achieve the desired cylinder end shortening. An example minimisation convergence plot is shown in Supplementary Note 8.

To simulate local probe experiments, in addition to fixing the end caps we also fix the position of a single node in the centre of the cylinder (thus mimicking a point probe). This point is moved radially inwards by a small increment and the free energy minimised. This increment-minimisation procedure is repeated until the entire pathway from the unbuckled state to a second minimum has been obtained.

In the free energy minimum survey, we access the many different dimpled states by performing a basin hopping step prior to each minimisation \cite{Wales2003}. To perform this step, we begin with the unbuckled cylinder, and make a random number of trial dimples to the initial node coordinates. Each trial dimple consists of a paraboloidic indentation radially into the cylinder, in which the indentation depth is allowed to vary up to $R_0/2$.  

The minimum energy pathways (MEPs) between any two minima of equal end shortening are found using the string method \cite{E2007}, which we augment for use with high-dimensional systems. To begin with, the end points are maximally aligned through rotation and reflection of the displacement fields. An initial string of 30 images is then formed which interpolates the coordinates of the two end points. One iteration of the algorithm consists of evolving each image in the downhill direction, then re-interpolating the images along the new string. The Euler and Runge-Kutta methods used in \cite{E2007} are however highly inefficient for the high-dimensional energy landscape considered here. Instead, we use 300 L-BFGS steps to rapidly converge the string to the MEP. A simple linear re-interpolation scheme is used, with the image density concentrated at the highest energy points along the string. This process is iterated until the $E_{\rm{r}}$ of the highest energy point along the string changes by less than $10^{-6}$ from the previous iteration. If intermediate minima exist along the pathway, a separate string is evolved for each, such that each pathway connects two minima via a single transition state. The Euler method is employed in the final stage to fine-tune the pathway, such that convergence is achieved when the RMS distance between the strings is less than $10^{-6}$. The transition state is then fine-tuned using the climbing string method with Euler steps \cite{Ren2013}, finishing once the RMS gradient is reduced below $10^-5$. Repeating the string algorithm to connect multiple end points forms a network of connected minima.

In order to show the general validity of this model, we further apply it to analyze the energy landscapes of the buckling of spherical caps in Supplementary Note 9. Our model has similar accuracy as the finite element model implemented in ABAQUS, and successfully captures the stable axisymmetrically inverted configuration of the spherical cap \cite{Taffetani2018}. The analysis is suitably rich that we reserve further discussion for another publication.

\bibliography{Bibliography}

\section*{Acknowledgements}
H.K. would like to acknowledge EPSRC for funding, Grant No. EP/P007139/1.

\section*{Data availability}
The datasets generated during and/or analysed during the current study are available from the corresponding author on reasonable request.

\section*{Code availability}
The lattice-spring energy landscape methods used in the current study are available from the corresponding author on reasonable request.

\section*{Competing interests}
The authors declare no competing interests.

\section*{Author contributions}
H.K. and T.Z. conceived the idea, designed the research and supervised the project. J.R.P. developed the energy landscape code, and performed simulations for exploring and controlling the buckling landscapes of thin shells. J.C. developed the triangulated lattice model, benchmarked it against ABAQUS, and performed simulations for local probe tests of shells. H.K. and J.R.P. drafted the manuscript and all authors contributed to the writing of the manuscript.

\end{document}


\twocolumn[
\begin{@twocolumnfalse}
\maketitle
\end{@twocolumnfalse}
]

\section*{Supplementary note 1: Single dimple stability}
\subsection*{Global free energy minima}

\begin{figure}[!ht]
\centering
\includegraphics[width=0.5\textwidth]{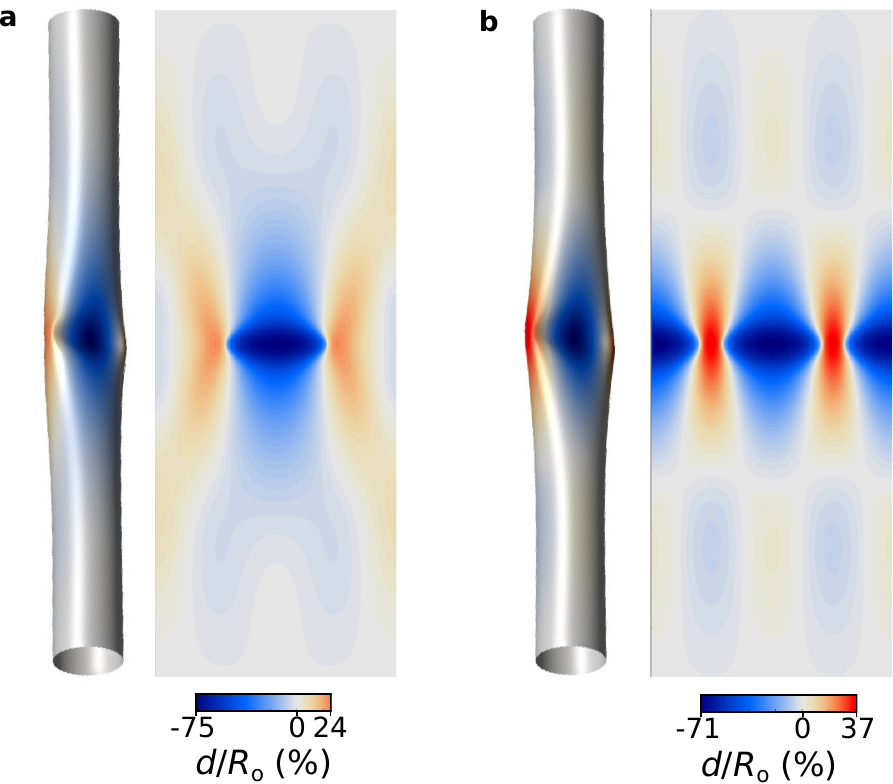}
\caption{Free energy minima 3D visualisations and radial displacement maps for a cylinder of $A_0$=10, $\lambda$=0.990, and an elastic control ratio = 2.5$\times10^3$. \textbf{a} Single dimple. \textbf{b} Two dimples.}
\label{fig:S1_dimp_stabil}
\end{figure}

Across the range of aspect ratios $A_0$ and shortening ratios $\lambda$, it was observed in Fig. 1d (main text) that the singly dimpled state was never the global free energy minimum. For cylinders of low aspect ratio, this observation is explained by the high localisation in both the axial and azimuthal directions, such that multiples dimples may form on the structure which have sufficient spacial separation to not interact. If a single dimple reduces the overall free energy of the system by exchanging stretching contributions for bending contributions, then a second spatially separated dimple reduces the free energy further by an equal amount. Thus, for any system with physical or elastic properties that enable multiple non-interacting dimples to form, the single dimple may never be the global energy minimum. 

We next consider the alternative case where multiple dimples interact strongly, an extreme example of which is illustrated in Supplementary Figure \ref{fig:S1_dimp_stabil}. As will be shown in Supplementary Figure \ref{fig:S2_self_int} and Supplementary Figure \ref{fig:S3_k10}, spatial localisation of a dimple is reduced on increasing $A_0$ and reducing the elastic control ratio, allowing for multiple dimples to interact over the shell. We therefore aim to probe the single dimple stability at extreme values of $A_0$ and elastic control ratios. The limit of this investigation is shown in Supplementary Figure \ref{fig:S1_dimp_stabil}, where $A_0$=10, $\lambda$=0.990, and the elastic control ratio = 2.5$\times10^3$. For reference, a material with Poisson ratio = 0.3 and elastic control ratio = 2.5$\times10^3$ yields a shell with radius to thickness ratio $R_0/t$=25. The singly dimpled state shown in Supplementary Figure \ref{fig:S1_dimp_stabil}a exhibits extreme deformation, with a maximum radial displacement of 75\% of $R_0$, and a reduced free energy of $E_r$=34.8. The free energy is still reduced on forming a second dimple however, shown in Supplementary Figure \ref{fig:S1_dimp_stabil}b, where $E_r$=31.9. Overall, over a large range of physically relevant conditions, the singly dimpled state is never the global free energy minimum.

\subsection*{Self-interaction}
\begin{figure}[!ht]
\centering
\includegraphics[width=0.5\textwidth]{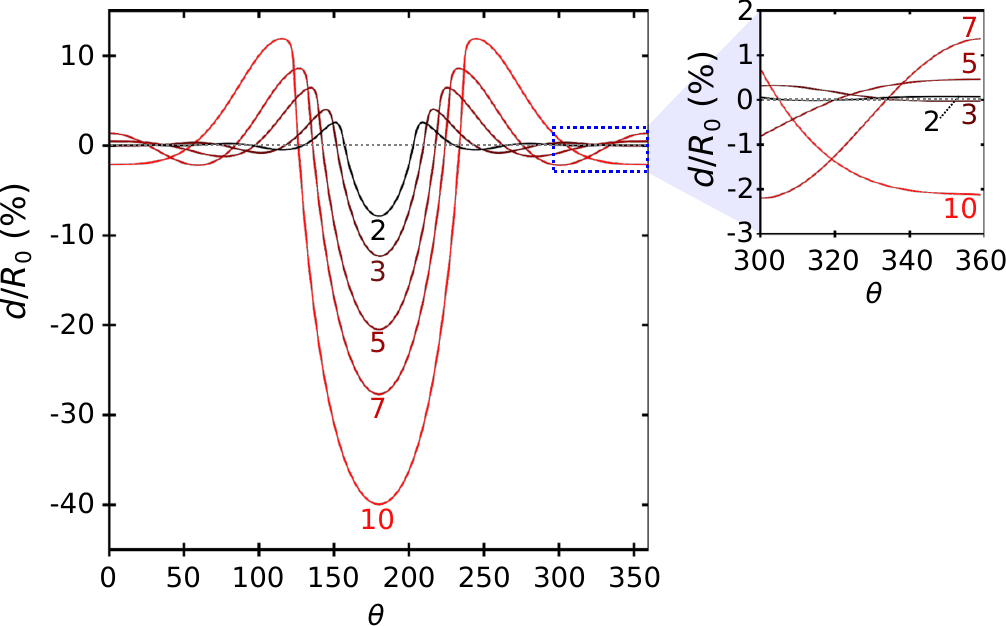}
\caption{Radial deformation profiles at the large-$\lambda$ stability limit of the the singly dimpled states, taken about the centre of the cylinder. Each profile is labelled with the corresponding $A_0$, and for each the elastic control ratio is 2.5$\times10^5$. Inset: a magnification of the deformation profile far from the dimple centre at $\theta=180^\circ$.}
\label{fig:S2_self_int}
\end{figure}

The low-$\lambda$ single dimple stability limit, shown in Fig. 1d (main text), exhibits a non-monotonic variation with $A_0$ when $A_0\ge$5. In Supplementary Figure \ref{fig:S2_self_int}, we plot the radial displacement of the single dimple about the centre of the cylinder at the low-$\lambda$ stability limit. It is observed that for $A_0$=2, the dimple is highly localised in the azimuthal direction, and does not extend around the full circumference. This is shown in Supplementary Figure \ref{fig:S2_self_int}(inset) as the radial deformation tending to zero far from the dimple centre.  This angular localisation decreases however as $A_0$ is increased, so that for $A_0\ge$5, the radial displacement far from the dimple centre is of the order of 1\% of $R_0$. Thus, self-interaction effects modify the single dimple stability for $A_0\ge$5, leading to the non-monotonic low-$\lambda$ stability limit on increasing  $A_0$.

\section*{Supplementary note 2: Further analyses of the local probe technique}

\subsection*{Accuracy of the unbuckled-single dimple local probe}

In Tables \ref{table:probe_acc_comp} and \ref{table:probe_acc_asp}, we compare the transition state energy $E_{\rm{TS}}$ from the MEP, with that obtained via the local probe technique for the unbuckled-single dimpled transition. In every case tested, the local probe achieves highly accurate estimates of the MEP transition state energy with a small percentage difference between the two (typically of the order of $10^{-3}$\%).  This is consistent with the typical accuracy of the local probe: successive increments change the energy by $10^{-5} - 10^{-4}$ close to the transition state. 

\begin{table}
\normalsize
\begin{adjustbox}{width=0.5\textwidth}
	\begin{tabular}{c  c  c  c}
	\hline
	$\lambda$ & $E_{\rm{TS}}$(MEP) & $E_{\rm{TS}}$(Probe) & Difference / \%  \\ \hline
	0.9990 & 7.42679	 & 7.42709	& 0.00399 \\
	0.9989 & 8.96955	 & 8.96976	& 0.00236 \\
	0.9988 &	 10.66295 &	10.66329	 & 0.00315 \\
	0.9987 &	 12.50415 &	12.50452	 & 0.00290 \\
	0.9986 &	 14.49440 &	14.49477	 & 0.00257 \\
	0.9985 &	 16.63210 &	16.63242	 & 0.00191 \\
	\hline
	\end{tabular}
	\end{adjustbox}
\caption{Comparision of the unbuckled-single dimpled transition state energy $E_{\rm{TS}}$(MEP) for the MEP and local probe technique at different compression ratios $\lambda$. All other physical and elastic parameters remain fixed:  $A_0$ = 2.0, $k_{\rm{stretch}}R_0^2/k_{\rm{bend}} = 2.5 \times 10^5$.}
\label{table:probe_acc_comp}
\end{table}

\begin{table}
\normalsize
\begin{adjustbox}{width=0.5\textwidth}
	\begin{tabular}{c  c  c  c}
	\hline
	$A_0$ & $E_{\rm{TS}}$(MEP) & $E_{\rm{TS}}$(Probe) & Difference / \%  \\ \hline
	0.8 &	5.75967 &	5.75968 &	0.00007 \\
	1.0 &	7.25528 &	7.25555 &	0.00374 \\
	2.0 &	14.49440	 & 14.49477 &	0.00257 \\
	3.0 &	11.09301	 & 11.09242 &	-0.00524 \\
	\hline
	\end{tabular}
	\end{adjustbox}
\caption{Comparision of the unbuckled-single dimpled transition state energy $E_{\rm{TS}}$(MEP) for the MEP and local probe technique at different aspect ratios $A_0$. All other physical and elastic parameters remain fixed:  $\lambda$ = 0.9986, $k_{\rm{stretch}}R_0^2/k_{\rm{bend}} = 2.5 \times 10^5$.}
\label{table:probe_acc_asp}
\end{table}

\subsection*{Efficacy of multiple local probes}

\begin{figure*}[p]
\centering
\includegraphics[width=0.85\textwidth]{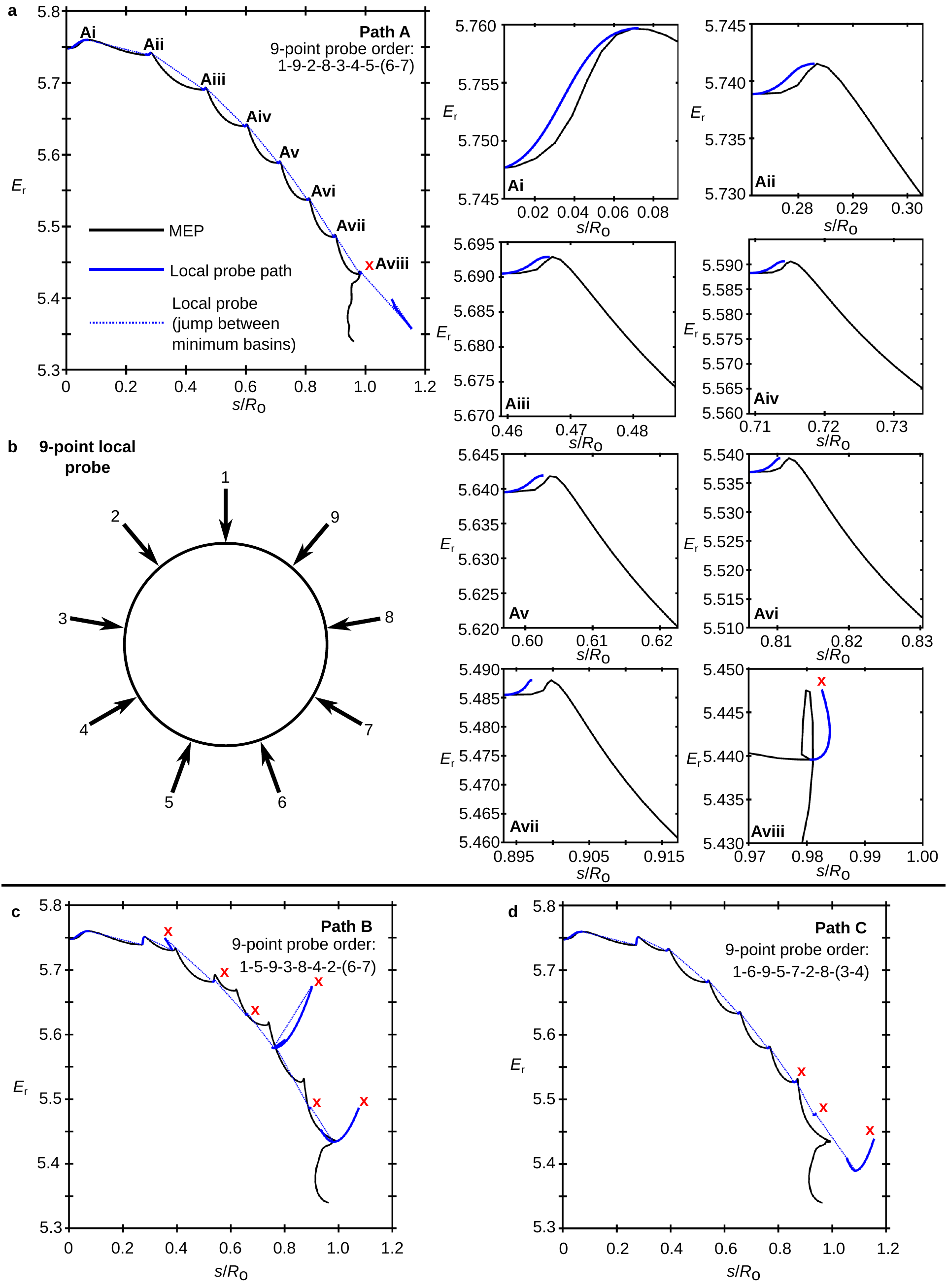}
\caption{Investigation of the accuracy of multiple local probing at recovering the MEP. \textbf{a} Path A: the multi-step unbuckled-(1$\times$9) global minimum energy pathway (black), compared to the nine-point local probing technique (blue). For the probing technique, discontinuous jumps between local basins of attraction are shown as dashed lines. Magnifications of each of the eight dimpling events are shown in \textbf{Ai}-\textbf{Aviii}. Red crosses indicate where the probing transition does not correspond to the transition in the MEP. \textbf{b} Labelling scheme of the nine local probes. \textbf{c} Nine-point local probing of Path B, showing early failure. \textbf{c} Nine-point local probing of Path C, showing late failure. For all paths, the path distance $s$ measures the distance in the lattice points from the current state to the unbuckled state. The order of the dimpling event is also shown for paths A, B, and C, with parenthesised locations indicating both dimples occur in the same transition. }
\label{fig:S4_9probe}
\end{figure*}

We have shown that the unbuckled-single dimple minimum energy barrier can be accurately accessed through the local probing technique for $A_0=0.8$, $\lambda=0.9986$. In Supplementary Figure \ref{fig:S4_9probe}, we investigate whether local probing can access more complex pathways. This is tested via attempting to reproduce the sets of transitions which occur in the multi-step unbuckled-(1$\times$9) pathways. This is attempted for the selected paths A, B, and C. We focus on Path A initially, reproduce in Supplementary Figure \ref{fig:S4_9probe}a.

To begin the local probe setup, we must estimate where the dimples will form without prior knowledge of the MEP. It is reasonable to assume the (1$\times$9) state is formed by nine separate dimpling transitions. which are equally spaced around the circumference of the cylinder centre. These locations are labelled in Supplementary Figure \ref{fig:S4_9probe}b. A number of choices exist with how to proceed with locally probing at nine locations, regarding in particular: whether the probes are applied individually or in unison, the depth each probe should be tested to, and whether probes should be removed or remain in place after a dimpling event. Here, we choose to test the pathway as simply as possible, in which: (1) A single probe is initiated at radius $R$ = $R_O$; (2) this probe is incremented radially inwards, up to the point where the energy of the system begins to reduce; (3) the system is then relaxed with the probe removed to form a new (dimpled) equilibrium configuration; (4) The next single probe is applied, and steps (1-3) repeated. By selecting the order of the probes, we may access different dimpling pathways. To compare as closely as possible to the MEPs here, we observe the dimpling order in paths A, B, and C, and perform the local probes in this order. The dimpling order is shown for paths A, B, and C, in Supplementary Figure \ref{fig:S4_9probe}a,c, and d respectively. The parenthesised pair of dimples in these orders indicate both dimples form in a single transition, but in general this would not be known \textit{a priori}. 

For path A in Supplementary Figure \ref{fig:S4_9probe}a, each of the eight separate transitions are labelled \textbf{Ai}-\textbf{Aviii}, and shown individually magnified. It can be seen that transition \textbf{Ai} is well reproduced in both energy and location by the local probe. As the series of transitions proceeds up to \textbf{Avii} however, although the local probe is able to accurately capture the transition state energy, the transition state location is obtained with decreasing accuracy. This can be rationalised by observing the movie of path A in Supplementary Movie 1. It can be seen that the dimples shift slightly in their location as the transition proceeds, which cannot be accommodated for in the fixed local probe simulation. In \textbf{Aviii}, the local probing technique fails to capture the MEP, as two dimples form in the same transition in the MEP, but not the probing simulation. As can be seen at the end of Supplementary Movie 1, the start of the transition sees one dimple partially form, then the other grows to match it, before both grow together to complete the transition.

For path B in Supplementary Figure \ref{fig:S4_9probe}c, the nine-point probing scheme departs from the MEP after only two transitions. This is because for this sequence of dimples, a probe applied at one point causes a large shift in the previously-formed dimple locations. Successive probing therefore continues to deviate further from the MEP. The same observation is made later in the pathway for path C in Supplementary Figure \ref{fig:S4_9probe}d.

Overall, this multiple local probing scheme is able to closely estimate the minimum energy barrier, but only when: (1) The dimpling event is single-dimpled; (2) the probe does not cause a large shift in previously-formed dimples; (3) the dimple formed by the probe does not undergo a large shift on removal of the probe. This latter point would be particularly challenging to maintain for irregularly dimpled structures, where a best estimate of the probe location may not be readily available.

\section*{Supplementary note 3: Elastic property effects on the free energy landscape}
\begin{figure}[!ht]
\centering
\includegraphics[width=0.5\textwidth]{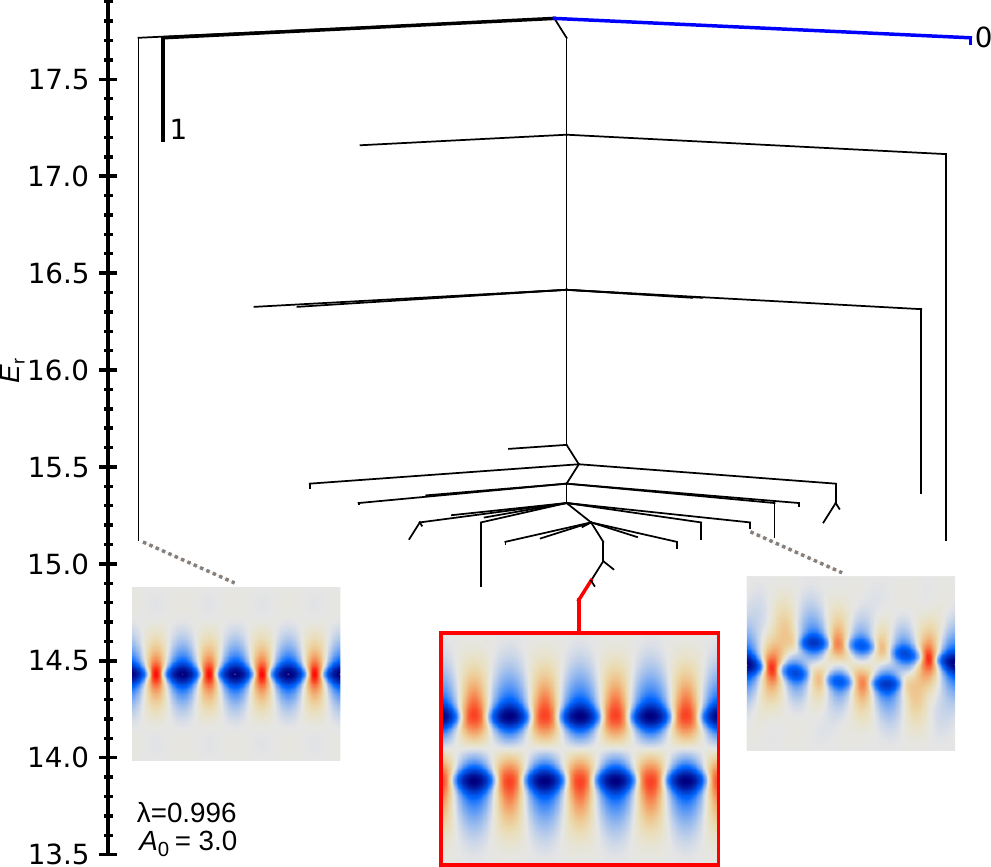}
\caption{Disconnectivity graph showing the minimum energetic barrier between any pair of states. The unbuckled and global minimum branches are coloured in blue (labelled '0') and red respectively. The global minimum radial displacement field is outlined in red. Representative examples of other minima are also shown. $A_0$ = 3.0, $\lambda$ = 0.996, elastic control ratio to 2.5$\times10^4$. The single dimple branch is labelled '1'.}
\label{fig:S3_k10}
\end{figure}

In Fig. 3c (main text), it was observed that for cylinders of high aspect ratio and thin shell thickness, at compression ratios where the unbuckled, singly dimpled and multiply dimpled states coexisted, the energy landscape became highly complex. This was observed for the cylinder of aspect ratio $A_0$ = 3.0, elastic control ratio 2.5$\times10^5$, and compression ratio $\lambda$=0.999. This complexity was manifest both in the number of minima, $\mathcal{O}(10^3)$, and multiple energy scales over which the landscape was 'rough'. Here, we observe the effect of decreasing the elastic control ratio (equivalent to increasing the shell thickness) on the landscape complexity. In Supplementary Figure \ref{fig:S3_k10}, we illustrate the energy landscape for a cylinder of the same aspect ratio, $A_0$ = 3.0, and use a compression ratio where the unbuckled, singly dimpled and multiply dimpled states also coexist, $\lambda$ = 0.996, but decrease the elastic control ratio to 2.5$\times10^4$. This decrease in the elastic control ratio, equivalent to a thickness increase of $\approx 3\times$, has a marked, simplifying affect on the energy landscape. The landscape is transformed from being rough and glassy ($\mathcal{O}(10^3)$ minima, bin population variance = 93), to being funnel-shaped with several deep states($\mathcal{O}(10^1)$ minima, bin population variance = 0.9). The global minimum is a ($2\times4$) diamond pattern, but a deep ($1\times4$) state also exists. At low energies the landscape is dominated by systems with 7 to 8 dimples. 

The dominating reason for this pronounced simplification is due to the size of each dimple relative to the cylinder area: whereas 8 dimples pack round the cylinder at the global minimum here, when the elastic control ratio was 2.5$\times10^5$, 12 were able to do so. The number of dimple arrangements possible on the cylinder here is therefore markedly reduced. It is also instructive to compare Fig. 3b (main text) where $A_0$ = 0.8, elastic control ratio 2.5$\times10^5$, and compression ratio $\lambda$=0.998, with the disconnectivity graph here. In Fig. 3b (main text), the strong confinement introduced by the dimples interacting with the fixed ends caused numerous deep states to exist, as well as for the majority of states to occur in the low-energy region of the landscape. Both of these observations are also made here in Supplementary Figure \ref{fig:S3_k10}, suggesting that by decreasing the elastic control ratio, the dimples are confined by the fixed ends. This is in contrast to the cylinder with $A_0$ = 3.0, elastic control ratio 2.5$\times10^5$, in which the dimples were weakly confined, giving rise to the glassy landscape.

\section*{Supplementary note 4: Minimum and barrier distributions in the free energy landscapes}
\begin{figure*}[!ht]
\centering
\includegraphics[width=\textwidth]{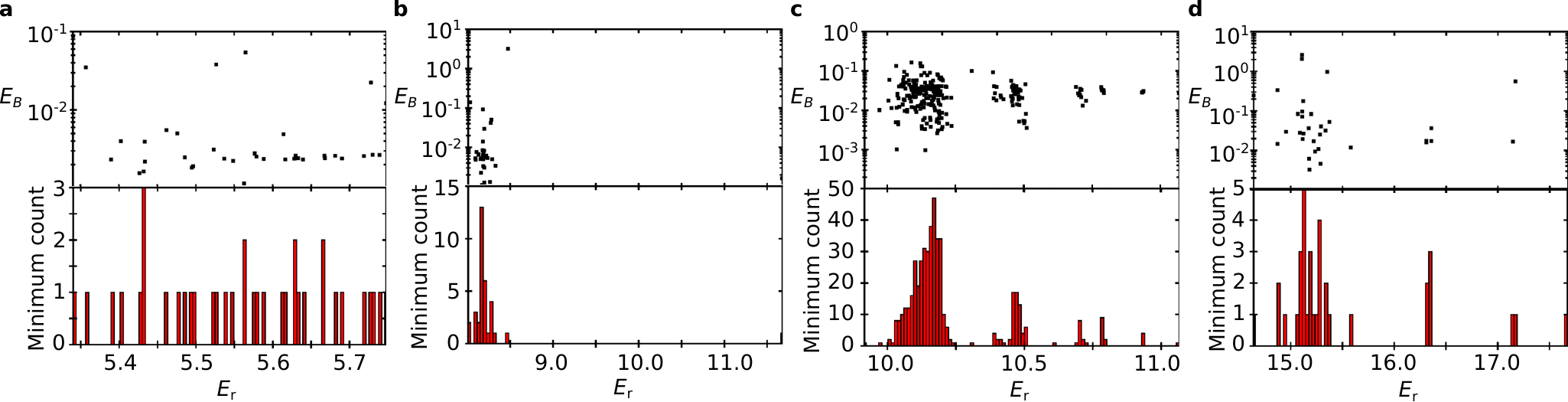}
\caption{Distributions of minimum energy barriers with stable state energies (top panels), and histograms of state frequency (lower panels) when states are counted within 100 bins over the landscape energy range. \textbf{a} $A_0$ = 0.8, $\lambda$=0.9986, elastic control ratio 2.5$\times10^5$. \textbf{b} $A_0$ = 0.8, $\lambda$=0.998, elastic control ratio 2.5$\times10^5$. \textbf{c} $A_0$ = 3.0, $\lambda$=0.999, elastic control ratio 2.5$\times10^5$. \textbf{d} $A_0$ = 3.0, $\lambda$=0.996, elastic control ratio 2.5$\times10^4$.}
\label{fig:S4_dist}
\end{figure*}

Throughout our disconnectivity graph discussions, two key properties have become useful comparative metrics between different buckling landscapes: the range of minimum energy barrier heights, and the uniformity of minimum distributions. In Supplementary Figure \ref{fig:S4_dist} (upper panels), we begin by plotting the minimum energy barrier height as a function of the energy of each state. The range in barrier energies are shown in Table \ref{table:metrics}. The deep states exhibited by the $A_0$ = 0.8, $\lambda$=0.998; and $A_0$ = 3.0, $\lambda$=0.996 systems mean the range of barrier energies for both span approximately 3 decades. This is compared to the 2 decade range in energy barriers for the systems lacking deep states: $A_0$ = 0.8, $\lambda$=0.9986; and $A_0$ = 3.0, $\lambda$=0.999.

Next, we partition each landscape  minimum energy range into 100 bins of equal width, and count the associated number of states within each. These state density distributions are plotted in Supplementary Figure \ref{fig:S4_dist} (lower panels). Uniquely in our study, for each dimpling configuration in the $A_0$ = 3.0, $\lambda$=0.999 system, a number of similarly structured minima exist with small energy barriers to their interconversion. For this reason, we clustered states with the same number of dimples and energy barrier less than $10^{-3}$. The uniformity of each distribution is quantified by calculating the variance in the bin populations, listed in Table \ref{table:metrics}. The glassy landscape of the $A_0$ = 3.0, $\lambda$=0.999 system is readily distinguishable from the large population variance, 93. The two roughness scales of the energy landscape also become apparent here: within clusters, barriers of less than $10^{-3}$ are present, associated with local movement of dimples in the same configuration, whereas the typical energy barrier between clusters is shown in Supplementary Figure \ref{fig:S4_dist}c to be of the order of $10^{-2}$.

\begin{table}
\normalsize
\begin{adjustbox}{width=0.5\textwidth}
	\begin{tabular}{c  c  c  c  c }

	\hline
	$A_0$ & $\lambda$ & Control ratio & Barrier range (decades) & Bin count variance  \\ \hline
	0.8 & 0.9986 & 2.5$\times10^5$ & 1.87 & 0.38 \\
	0.8 & 0.998 & 2.5$\times10^5$ & 3.46 & 2.1 \\	
	$^*$3.0 & 0.999 & 2.5$\times10^5$ & 2.24 & 93 \\
	3.0 & 0.996 & 2.5$\times10^4$ & 2.90 & 0.93 \\	
	\hline
	
	\end{tabular}
	\end{adjustbox}
\caption{Comparative metrics for the four cylindrical systems reviewed. $^*$Analysis performed after clustering.}
\label{table:metrics}
\end{table}

\section*{Supplementary note 5: Landscape biasing for targeted design}
\subsection*{General workflow}
\begin{figure}[!ht]
\centering
\includegraphics[width=0.5\textwidth]{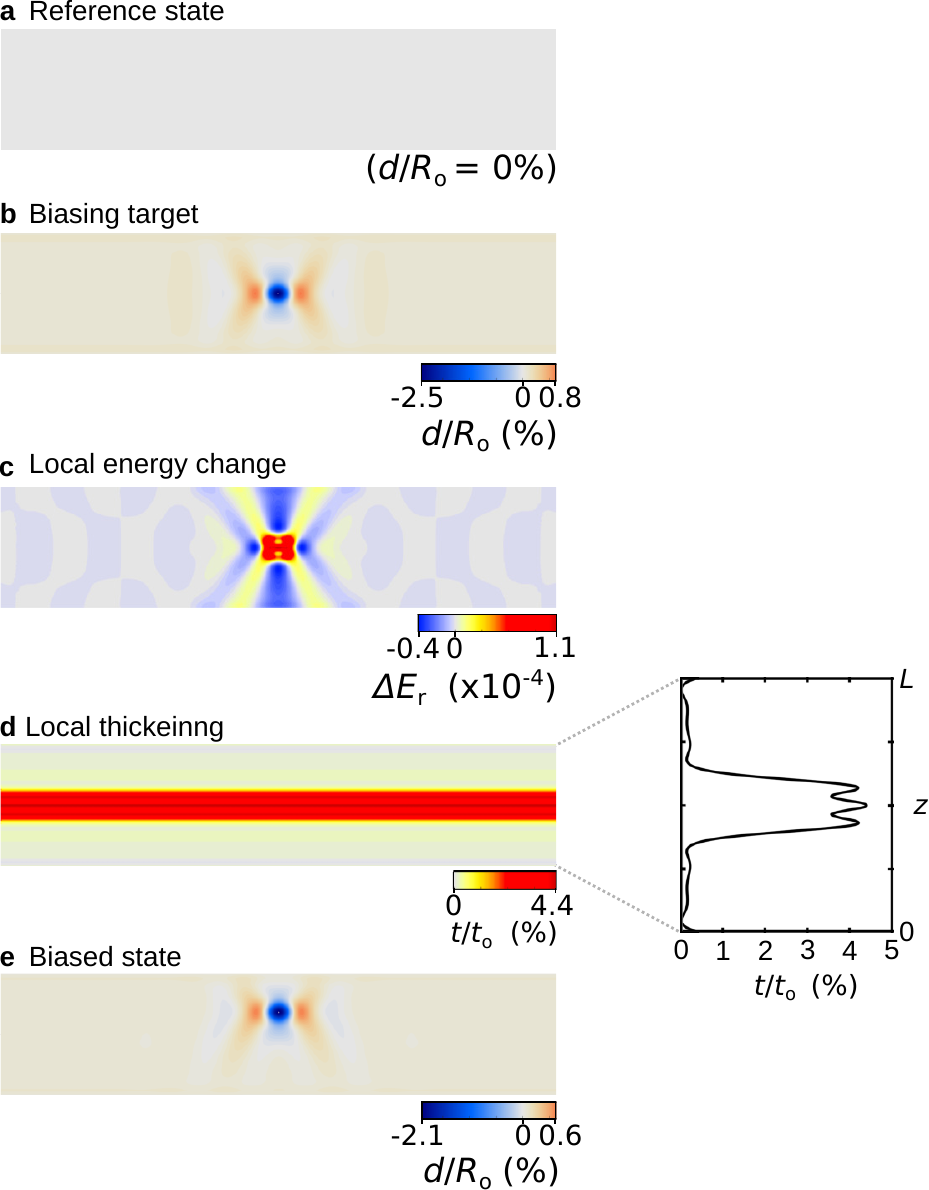}
\caption{Sequential workflow for the landscape biasing process to design the buckling responses.}
\label{fig:S5_bias}
\end{figure}

The sequence of steps required for landscape biasing is shown in Supplementary Figure \ref{fig:S5_bias}. To begin with, a reference state is chosen which undergoes the target buckling response. The example reference shown in Supplementary Figure \ref{fig:S5_bias}a is the unbuckled state for the $A_0$ = 0.8, $\lambda$=0.9986 system, with the example biasing target shown as the 0-1 dimple transition state in Supplementary Figure \ref{fig:S5_bias}b. 

Next, the elastic potential energy associated with each node, i, in the triangulated mesh is calculated for both the biasing target, $E_{\rm{target}}(i)$, and the reference $E_{\rm{ref}}(i)$. The difference,
\begin{equation}
\Delta E(i)=E_{\rm{target}}(i)-E_{\rm{ref}}(i),
\end{equation}
is shown for the biasing example in Supplementary Figure \ref{fig:S5_bias}c.

Numerous choices now exist in using the local energy difference to inform the local thickening. We now demonstrate two example methods: one was used to bias against the 0-1 transition state, and the other for the (1$\times$8) minimum.

To bias against the 0-1 transition state, $\Delta E(i)$ is shifted by the minimum value of $\lbrace\Delta E(i)\rbrace$, then normalised, so that the weighting of element i is expressed:
\begin{equation}
w_i=\frac{\Delta E(i)-\rm{min}\lbrace\Delta E(i)\rbrace}{\sum_i\left[\Delta E(i)-\rm{min}\lbrace\Delta E(i)\rbrace\right]}.
\end{equation}
The local weights are then used to simulate a local thickening of the elastic shell, so that the ratio of the new thickness of element i, $t_i$, to the local thickness of the reference state, $t_i^o$, is
\begin{equation}
\frac{t_i}{t_i^o}=1+\alpha w_i,
\end{equation}
where $\alpha$ is the biasing amplitude, and represents the prescribed fractional increase in mass of the shell, if we make the reasonable assumption that the mass increases in proportion to the simulated shell thickness. For the 0-1 transition state example, we make a preliminary step in which the $\Delta E(i)$ are averaged about the circumference of the cylinder. The resultant local thickening of this treatment, using a biasing amplitude $\alpha = 0.01$, yields the thickening map illustrated in Supplementary Figure \ref{fig:S5_bias}d. The inset shows the z-variation in the local thickening profile.

To bias for the (1$\times$8) minimum, the local weighting $w_i$ is obtained via shifting $\Delta E(i)$ with respect to the average value of $\lbrace\Delta E(i)\rbrace$, and rescaled by the maximum value of $\lbrace|\Delta E(i)|\rbrace$, such that,
\begin{equation}
w_i=\frac{\Delta E(i)-\rm{average}\lbrace\Delta E(i)\rbrace}{\rm{max}\lbrace|\Delta E(i)|\rbrace}.
\end{equation}
The local thickening ratio, $t_i/t_o$ is then expressed,
\begin{equation}
\frac{t_i}{t_i^o}=1-\alpha w_i.
\end{equation}
This treatment ensures that the total mass increase of the biased cylinder is zero, as $\sum_i w_i=0$, and the magnitude of the greatest local fractional change in thickness is the biasing amplitude $\alpha$.

The final step is to transform the local elastic constants to effectively simulate the local thickening. The new local stretching constant between nodes i and j, $k^{\rm{stretch}}_{\rm{ij}}$, relative to the local reference constant $k^{\rm{stretch}}_{\rm{o}}$ is
\begin{equation}
k^{\rm{stretch}}_{\rm{ij}} = \frac{k^{\rm{stretch}}_{\rm{o}}}{2} \left[ \frac{t_i}{t_i^o} + \frac{t_j}{t_j^o} \right].
\end{equation}
Similarly, the new local dihedral bending constant associated with nodes i, j, k, and l, $k^{\rm{bend}}_{\rm{ijkl}}$, relative to the local reference constant $k^{\rm{bend}}_{\rm{o}}$ is
\begin{equation}
k^{\rm{bend}}_{\rm{ijkl}} = \frac{k^{\rm{bend}}_{\rm{o}}}{4} \left[ \left(\frac{t_i}{t_i^o}\right)^3 + \left(\frac{t_j}{t_j^o}\right)^3 + \left(\frac{t_k}{t_k^o}\right)^3 + \left(\frac{t_l}{t_l^o}\right)^3 \right].
\end{equation}

The example result of the local thickening treatment for the 0-1 transition state is shown in Supplementary Figure \ref{fig:S5_bias}e: the transition state has been shifted off-centre, increasing the minimum energy barrier.

\subsection*{Interpolated biasing}

\begin{figure}[!ht]
\centering
\includegraphics[width=0.5\textwidth]{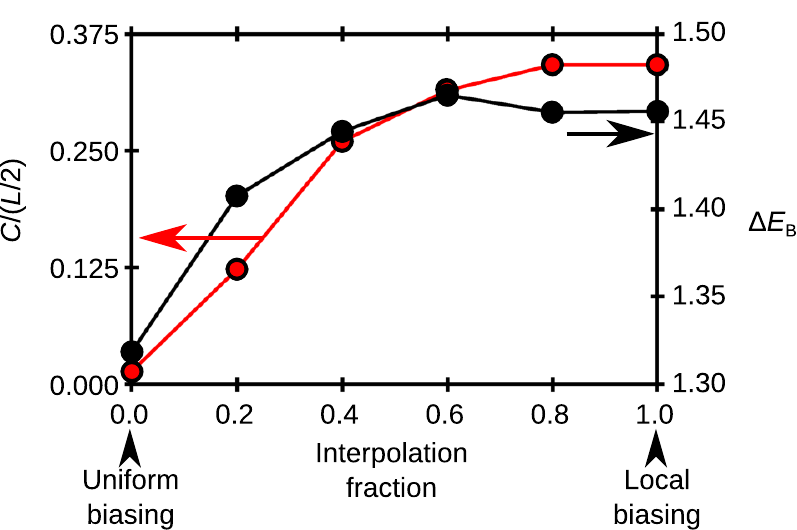}
\caption{Impact of the biasing localisation on the displacement of the dimple $C$ from the centre of the cylinder length (red data, left axis), and on the energy barrier $\Delta E_{\rm{B}}$ (black data, right axis). }
\label{fig:S7_bias_interp}
\end{figure}

Upon biasing against the unbuckled-single dimple transition state for the $A_0$=0.8, $\lambda$=0.9986 system, the biased transition state is observed to form off-centre. This displacement is caused by the energetic penalty associated with deforming the locally thickened central region of the cylinder. As it is also energetically unfavourable to deform the cylinder close to the simply supported ends, the dimple is expected to form at a location $C$ between 0 and $L/2$ from the centre of the cylinder. We test the hypothesis that symmetry breaking is caused by local thickening about the centre with the following set of simulations (all at the prescribed 1\% mass increase). We begin by testing the energy barrier $\Delta E_{\rm{B}}$ on the uniformly thickened cylinder, which we assign an 'interpolation fraction' of 0. We then obtain the energy barriers a series of cylinders with interpolated thickening fields between the uniform cylinder, and the fully biased cylinder (which we assign an interpolation fraction of 1). The dimple position and energy barrier of these interpolations are shown in Supplementary Figure \ref{fig:S7_bias_interp}.

As anticipated, the more localised the biasing is to the centre, the larger the energetic penalty for deforming the centre, so the further away the dimple forms. However, interestingly there is an optimum interpolation fraction between local and uniform, at approximately 0.6. We can rationalise this observation by considering that to maximise the barrier to dimpling, the interpolation fraction should be large enough to force the dimple off-centre, but not so large that the dimple forms in a region of low thickness.

\section*{Supplementary note 6: Resolution tests}

\begin{figure}[!ht]
\centering
\includegraphics[width=0.5\textwidth]{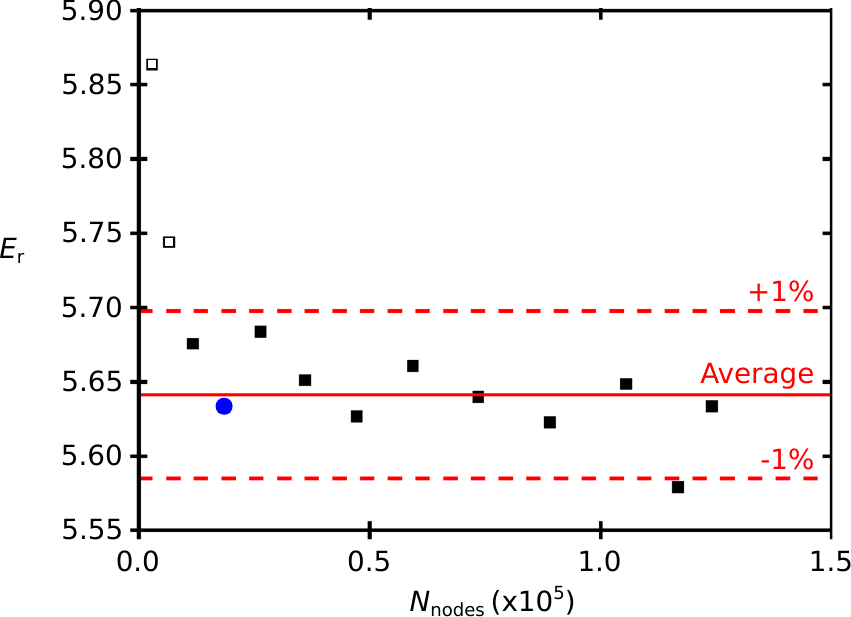}
\caption{Energy of a multiply dimpled state for $A_0$=0.8 and $\lambda$=0.9986, as computed with meshes using different numbers of nodes, $N_{\rm{nodes}}$. The average energy of the filled data is shown in as the red solid line, with $\pm$1\% margins shown as the red dashed lines. The $N_{\rm{nodes}}$ used in this work is plotted as the blue circle.}
\label{fig:S8_restest}
\end{figure} 

In Supplementary Figure \ref{fig:S8_restest}, the influence of the triangulated mesh node density on the computed energy is tested. To perform the test, a multiply dimpled state on the $A_0$=0.8, $\lambda$=0.9986 cylinder was chosen. The energy for this state was then calculated using different numbers of nodes, $N_{\rm{nodes}}$. Supplementary Figure \ref{fig:S8_restest} shows for $N_{\rm{nodes}}>10^4$ (filled data), the accuracy of the computed energy cannot be further improved. By averaging the energies of systems with $N_{\rm{nodes}}>10^4$, we show that all meshes produce energies consistent to within $\approx$1\% of this average. The resolution we choose to perform further simulations at is highlighted with a blue circle, balancing accuracy with computational efficiency.

\section*{Supplementary note 7: Model tests}

\subsection*{Model validation}
\begin{figure}[!ht]
\centering
\includegraphics[width=0.5\textwidth]{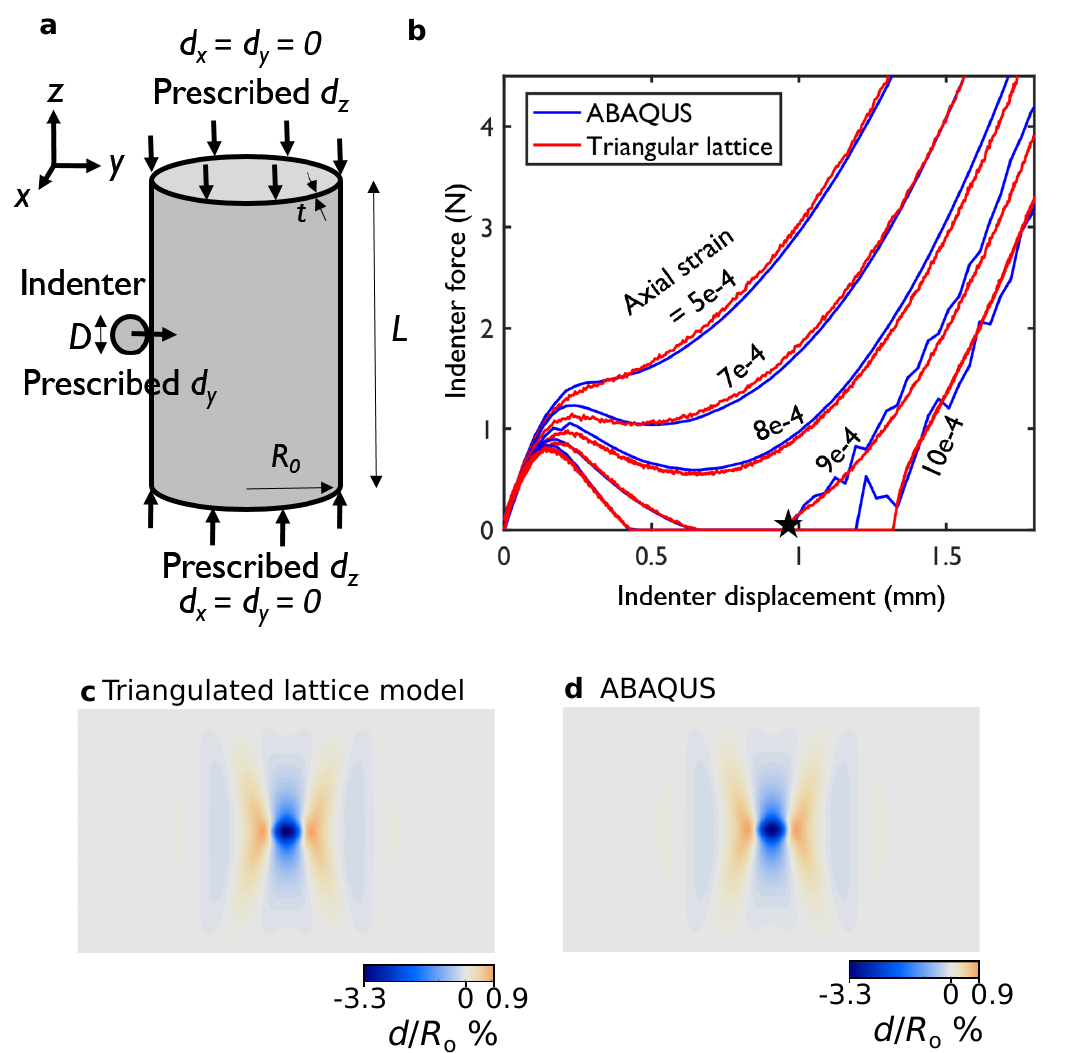}
\caption{Validation of the triangulated lattice model against ABAQUS. \textbf{a} Triangular lattice model for cylindrical shell. \textbf{b} Indenting force-displacement curves obtained from ABAQUS and triangular lattice simulations at different axial compressions. \textbf{c}, \textbf{d} Contour plots of the radial displacement fields at the local energy minima of the axial strain = $9\times{10}^{-4}$ curves (black star in b) }
\label{fig:S6_validation}
\end{figure}

To verify the triangular lattice model, we first conducted simulations of indenting a cylindrical shell under axial compression \cite{Virot2017} using the triangular lattice model and ABAQUS/Explicit \cite{ABAQUS2001}. The cylindrical shell has radius $R_0=28.6$ mm, length $L_0=107$ mm, and thickness $t$=0.104 mm. It is made of linear elastic material with Young's modulus $Y=71000$ N ${\rm mm}^{-2}$ and Poisson's ratio $\nu=0.3$. The indenter is a rigid sphere with diameter $D=4.7$ mm. The cylindrical shell is modelled with 4-node shell element (S4R) in ABAQUS. The shell is first compressed axially by assigning a constant velocity at the bottom and top edges. This corresponds to the "simply supported" boundary conditions for thin shells. After a targeted axial compressive strain is reached, the velocities of the top and bottom edges are set to zero. The indenter is then approached to the cylinder with a constant velocity perpendicular to the shell surface to generate local probe deformation. (Supplementary Figure \ref{fig:S6_validation}a). Different loading rates are checked to make sure it is a quasi-static process. The indenting force-displacement curves of the triangular lattice model and the shell element in ABAQUS are in agreement, as shown in Supplementary Figure \ref{fig:S6_validation}b. The displacement maps of the dimpled structures are also compared in Supplementary Figure \ref{fig:S6_validation}c,d, and are in agreement to within 2\%. Moreover, our simulations captured the instabilities during indentation, which have been reported in experiments \cite{Virot2017}.

\begin{figure}[!hb]
\centering
\includegraphics[width=0.5\textwidth]{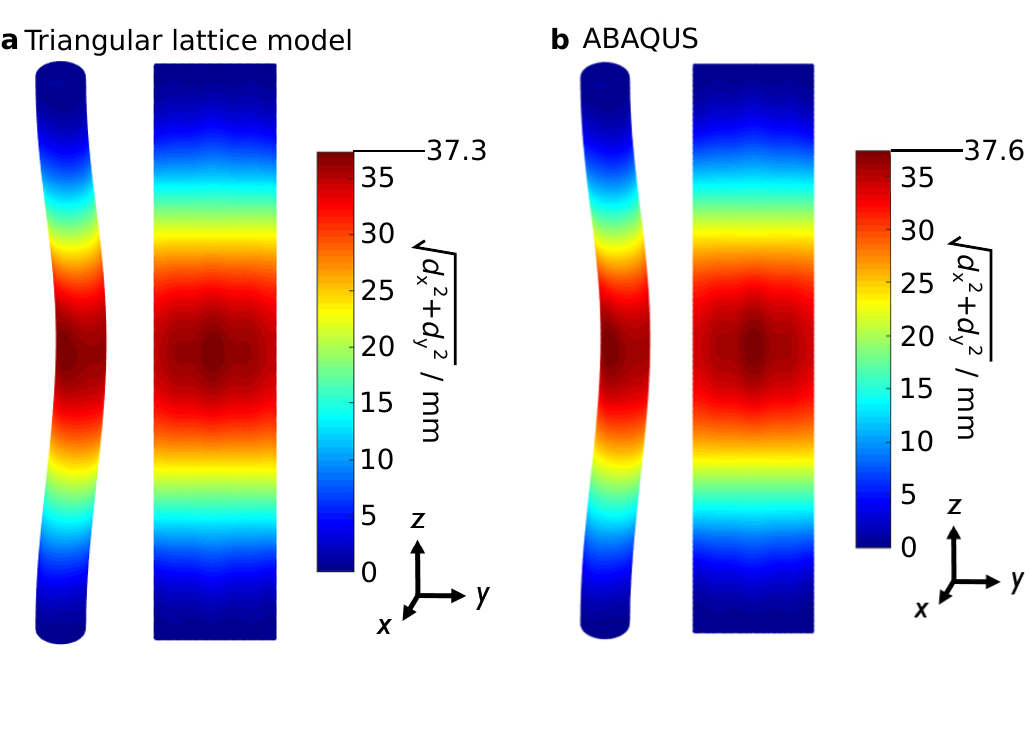}
\caption{Bending tests for a cylindrical shell of $R_0$=28.6 mm, $L_0$=858 mm, $t$=3.575 mm, $Y$=71000 N ${\rm mm}^{-2}$, and $\nu$=0.3. \textbf{a},\textbf{b} Contour plots of lateral deflection fields obtained from ABAQUS and triangular lattice simulations under the an axial strain of 0.024.}
\label{fig:S7_beam}
\end{figure}

We then simulated bending of a slender cylindrical shell under an axial compressive strain slightly above Euler's buckling load of a beam using the same boundary conditions. The comparison between our triangulated lattice model and ABAQUS is presented in Supplementary Figure \ref{fig:S7_beam}. Although there are no constraints for shell edge rotating, the bottom and top surfaces remain flat due to the finite size of the cross section. Therefore we need to apply the "clamped boundary" in the framework of beam theory. The parameters for cylindrical shell were chosen to be $R_0$=28.6 mm, $L_0$=858 mm, $t$=3.575 mm, $Y$=71000 N ${\rm mm}^{-2}$ and $\nu$=0.3. This makes sure its global (beam bending) buckling strain $\epsilon_{cr}^{\rm{global}}=\frac{\pi^2I}{{A\left(0.5L\right)}^2}\approx0.0220$ \cite{Bazant2010} is much smaller than its local (plate bending) buckling strain $\epsilon_{cr}^{\rm{local}}=\frac{t}{\sqrt{3(1-\nu^2)}R}\approx0.0756$ \cite{Hutchinson2018}, where $I$ and $A$ are the second area moment and area of cross section. The shell is first compressed until its axial strain reaches 0.024, slightly above $\epsilon_{cr}^{\rm{global}}$. After applying a small indenting force to initiate global buckling, the cylinder relaxes to an equilibrium post-buckling shape, shown in Supplementary Figure \ref{fig:S7_beam}a and b. The lateral deflection fields of both models are in agreement, as shown in \ref{fig:S7_beam}a and b, with a difference of less than 1\% in maximum values.

\section*{Supplementary note 8: L-BFGS convergence}

\begin{figure}[!t]
\centering
\includegraphics[width=0.5\textwidth]{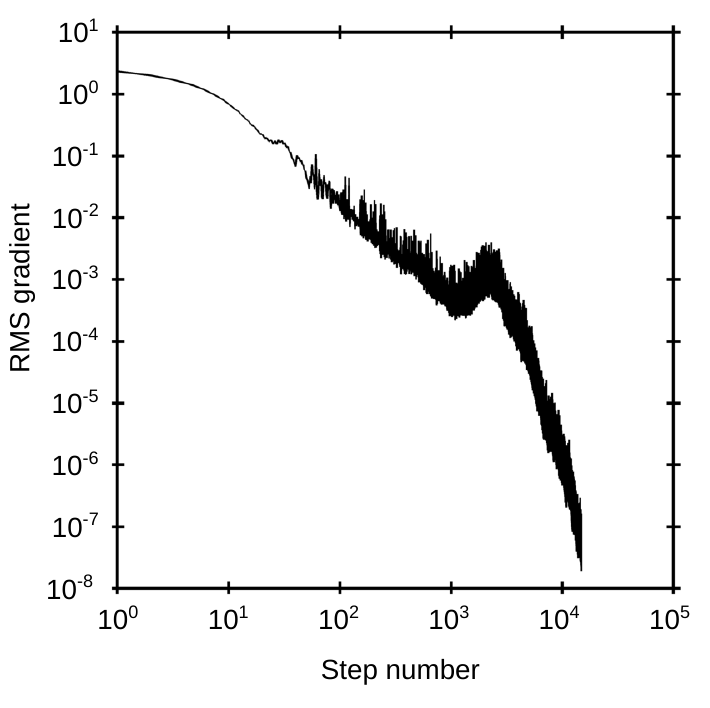}
\caption{Example convergence of a perturbed cylinder to a buckled state using the L-BFGS algorithm.}
\label{fig:S10_conv}
\end{figure}

We show an example convergence plot in Supplementary Figure \ref{fig:S10_conv} for the minimisation of a perturbed cylinder to a buckled state using the L-BFGS algorithm. Typically, convergence occurs when the energy change on successive steps is less than 10-11\% of the energy of a minimum.

\section*{Supplementary note 9: Generalising the model}

\begin{figure}[!ht]
\centering
\includegraphics[width=0.5\textwidth]{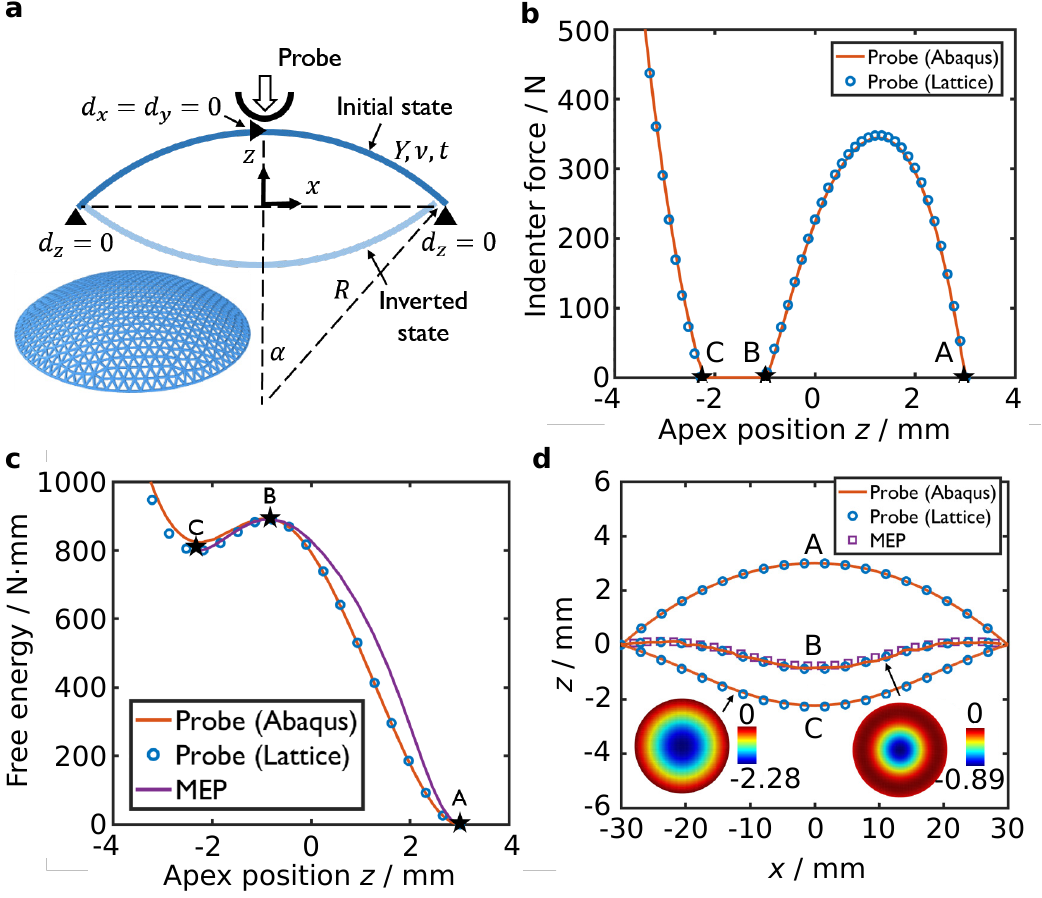}
\caption{\textbf{a} Parameters and boundary conditions in the spherical cap problem. The parameters chosen in our study are $Y$=71000 MPa, $\nu$=-0.3, $t$=0.7984 mm, $R$=151 mm and $\alpha$=0.2 rad. \textbf{b} Indenter force-apex position curves. \textbf{c} Free energy profile along the minimum energy path (MEP) connecting the initial state A and inverted state C and passing through the transition state B, compared with the free energy profile along the probe paths. \textbf{d} Deformation profiles of the cylindrical cap corresponding to the three states on the MEP. }
\label{fig:S9_sphere}
\end{figure} 

In the continuum limit, it has been shown that the effective bending energy of the perfect triangular lattice can be expressed as \cite{Schmidt2012,Wan2015},
\begin{equation}
E_{\rm{bending}}=\frac{k_{ref}^{\rm{bend}}}{4\sqrt3}\ \int{(3H^2-8K)dA},
\label{eq:sijc1}
\end{equation}
 where $H$ is the mean curvature, and $K$ is the Gaussian Curvature. For a continuum sheet, the bending energy can be written as \cite{Wan2015,Liang2009},
\begin{equation}
E_b=\ \int\left(\frac{1}{2}\kappa H^2+\kappa_GK\right)dA,                                             
\label{eq:sijc2}
\end{equation}  
where $\kappa=\left(\frac{Yt^3}{12\left(1-\nu^2\right)}\right)$ is the bending rigidity, and the Gaussian rigidity $\kappa_G=-\left(\frac{Yt^3}{12\left(1+\nu\right)}\right)$. $Y$ and $\nu$ denote Young's modulus and Poisson's ratio respectively.

By comparing Eqs. \eqref{eq:sijc1} and \eqref{eq:sijc2}, the effective Poisson's ratio of the triangular lattice is $\nu_{\rm{bending}}=-1/3$. It should be pointed out that the in-plane stretching energy and out-of-plane bending energy in the triangular lattice model are decoupled. Therefore the effective Poisson's ratio derived from bending energy has no effect on the Poisson's ratio in the in-plane deformation, which has been shown to be $\nu_{\rm{stretching}}$=1/3 \cite{Seung1988}.

To show our method can be generalized to other geometries, we consider the buckling of a spherical cap shell shown in Supplementary Figure \ref{fig:S9_sphere}a. The undeformed zero-energy shape of the shell is cut from a sphere with radius $R$, Young's modulus $Y$, Poisson's ratio $\nu$ and thickness $t$, with angle $\alpha$ between the pole and the edge, represented by the dark blue arc. Its edge is free to rotate but restricted to move within the $z$=0 plane ($u_z$=0), and its apex is fixed in the $x-y$ direction ($u_x$=$u_y$=0). For a spherical cap with large enough $R/t$, when probed with an indenter moving down along the $z$ axis, the shell will snap to an inverted local minimum, represented by the light blue arc, once the indenter displacement reaches a critical value \cite{Taffetani2018}.
 
For the triangular lattice model, we mesh the spherical cap with arbitrarily shaped triangles instead of equilateral triangles in cylindrical shells, as spherical shells cannot be isometrically mapped to simple planar shapes. The corresponding bending energy elastic constants for general triangulated meshes are \cite{Garg2007}:
\begin{equation}
k_{\alpha\beta}^{\rm{bend}}=\frac{l_{\alpha\beta\mathrm{\ }}^2}{A_\alpha+A_\beta}\frac{Yt^3}{12(1-\nu^2)}, 
\label{eqsijc3}
\end{equation}
where $A_\alpha$,$A_\beta$ are the triangular areas and $l_{\alpha\beta}$ is the length of the shared edge of two triangles. The equilibrium values of the dihedral angle $\theta_{\alpha\beta}^0$ are also computed based on the initial coordinates of the nodes on the spherical cap. 
To verify the results of triangular lattice model, we also conduct finite element (FE) simulations using shell elements in ABAQUS. It should also be pointed out that we take Poisson's ratio $\nu$ to be -0.3 in the FE model to match the ratio between Gaussian and bending rigidities in the triangular lattice model. This issue is not a concern in the main text as Gaussian rigidity does not play a significant role in the buckling of cylindrical shells, whose Gaussian curvature is zero. But in our preliminary indenting simulations of spherical caps, the force-displacement curves vary significantly after choosing different Poisson's ratios in the continuum model. In principle, the in-plane stretching energy will be also changed when varying Poisson's ratios in FE simulation, as the bending and stretching energy are coupled in the continuum model. However, the in-plane stiffness has little influence in the current study because buckling of spherical caps is dominated by bending energy. 
The parameters chosen in this study are $Y$=71000 N ${\rm mm}^{-2}$, $\nu=$-0.3, $t$=0.7984 mm, $R$=151 mm and $\alpha$=0.2 rad, so that the spherical cap has an inverted local minimum \cite{Taffetani2018}. We first simulate probing the shell with a hemispherical indenter modeled with triangular lattice and finite element in ABAQUS. Supplementary Figure \ref{fig:S9_sphere}b compares the indenter force – apex position $z_a$ curve of the two methods, and marks in black star the undeformed local minimum A, transition state B and inverted local minimum C. Supplementary Figure \ref{fig:S9_sphere}c compares the free energy profiles along the probe paths and the minimum energy path (MEP) connecting A and C obtained by string method. Similar to the unbuckled – single dimple transition of cylindrical shells, probing meets the MEP at the transition state B, but does not access the MEP generally. Supplementary Figure \ref{fig:S9_sphere}d compares the deformation profiles of the three states obtained by probing and string method. 
This preliminary work demonstrates that the methodology of integrating triangular lattice model and string method can capture the snap-through buckling of a spherical cap shell. Future work needs to be done to explore the energy barriers and MEP for more complicated buckling states, such as non-axisymmetric modes \cite{Taffetani2018}. In addition, the current triangular lattice has a fixed ratio between Gaussian and bending rigidities, corresponding to a Poisson's ratio $\nu$=-0.3. This is also understandable because there is only one parameter associated with bending in the current triangular lattice model, which cannot produce two different bending rigidities independently. It will be of great interests in developing a more general lattice model capable of describing materials with a wide range of Poisson's ratios.

\bibliography{Bibliography}